\documentclass[11pt,a4paper]{article}

\pdfoutput=1
\usepackage{graphicx}
\usepackage{jheppub}
\usepackage{multirow}
\usepackage{tabulary}
\usepackage{pifont}
\usepackage{float}
\usepackage{bbold}

\newcommand{\Mr}{\mathbf{M}_R}
\newcommand{\Yl}{\mathbf{Y}_\ell}
\newcommand{\Ynu}{\mathbf{Y}_\nu}
\newcommand{\Ytnu}{\widetilde{\mathbf{Y}}_\nu}
\newcommand{\Mnu}{\mathbf{M}_\nu}
\newcommand{\Ur}{\mathbf{U}_R}
\newcommand{\Unu}{\mathbf{U}_\nu}
\newcommand{\Ul}{\mathbf{U}_\ell}
\newcommand{\U}{\mathbf{U}}
\newcommand{\OCI}{\mathbf{O}}
\newcommand{\OI}{\textbf{O}_\text{I}}
\newcommand{\OII}{\textbf{O}_\text{II}}
\newcommand{\OIII}{\textbf{O}_\text{III}}
\newcommand{\dmatm}{\Delta m^2_{31}}
\newcommand{\dmsol}{\Delta m^2_{21}}
\newcommand{\Xnu}{\mathbf{X}_\nu}
\newcommand{\XN}{\mathbf{X}_R}
\newcommand{\Xhnu}{\widehat{\mathbf{X}}_\nu}
\newcommand{\XhN}{\widehat{\mathbf{X}}_R}

\newcommand{\cmark}{\ding{51}}
\newcommand{\xmark}{\ding{55}}

\newcolumntype{K}[1]{>{\centering\arraybackslash}m{#1}}
\newcommand\tsp{\rule{0pt}{0.49cm}}
\newcommand\bsp{\rule[-2.1ex]{0pt}{0pt}}
\newcommand\tspp{\rule{0pt}{0.58cm}}
\newcommand\bspp{\rule[-2.3ex]{0pt}{0pt}}
\newcommand{\specialcell}[2][c]{%
	\begin{tabular}[#1]{@{}c@{}}#2\end{tabular}}

\title{Combining texture zeros with a remnant CP symmetry in the minimal type-I seesaw}

\author[a]{D. M. Barreiros,}
\author[b,a]{R. G. Felipe}
\author[a]{and F. R. Joaquim}

\affiliation[a]{CFTP, Departamento de F\'{\i}sica, Instituto Superior T\'ecnico, Universidade de Lisboa,\\
	Avenida Rovisco Pais 1, 1049-001 Lisboa, Portugal}
\affiliation[b]{Instituto Superior de Engenharia de Lisboa,\\ 
	Rua Conselheiro Em\'{\i}dio Navarro 1, 1959-007 Lisboa, Portugal}

\emailAdd{debora.barreiros@tecnico.ulisboa.pt}
\emailAdd{filipe.joaquim@tecnico.ulisboa.pt}
\emailAdd{ricardo.felipe@tecnico.ulisboa.pt}

\abstract{In the framework of the two right-handed neutrino seesaw model, we consider maximally-restrictive texture-zero patterns for the lepton Yukawa coupling and mass matrices, together with the existence of a remnant CP symmetry. Under this premise, we find that several textures are compatible with the most recent data coming from neutrino oscillation and neutrinoless double beta decay experiments. It is shown that, the maximum number of allowed texture zeros in the Dirac Yukawa coupling matrix is two, for an inverted neutrino mass spectrum. In contrast, for Yukawa coupling matrices with just one texture zero, both normal and inverted orderings of neutrino masses are compatible with data. In all cases, the predictions for the low-energy Dirac and Majorana CP-violating phases, and for the effective mass parameter relevant in neutrinoless double-beta decay experiments, are presented and discussed. We also comment on the impact of future experimental improvements in scrutinising texture-zero patterns with a remnant CP symmetry, within the minimal version of the seesaw mechanism considered here.}


\arxivnumber{1810.05454}

\begin{document}
	
\maketitle
\flushbottom

\section{Introduction}
\label{sec1}
The observation of neutrino oscillations, which require the existence of neutrino masses and mixing, provides an irrefutable evidence for physics beyond the Standard Model (SM). Experimentally, three lepton mixing angles and two neutrino mass-squared differences have been precisely measured~\cite{deSalas:2017kay,Esteban:2016qun,Capozzi:2016rtj}. Also, hints for a nonzero Dirac CP-violating phase have been found~\cite{Abe:2018wpn,NOvA:2018gge}, being the best-fit value obtained from global analysis of neutrino oscillation data close to $3\pi/2$~\cite{deSalas:2017kay,Esteban:2016qun,Capozzi:2016rtj}. On the other hand, the Majorana character of neutrinos (which would imply the existence of Majorana phases) is still to be scrutinised, with the most promising efforts coming from searches for neutrinoless double-beta ($0\nu\beta\beta$) decays~\cite{DellOro:2016tmg,Vergados:2016hso,Giuliani:neutrino}. Furthermore, cosmological data and beta-decay experiments, indicate that neutrino masses are at least six orders of magnitude smaller than charged-lepton masses, suggesting that neutrinos may be particles of a different nature.

In the context of SM extensions, the seesaw mechanism~\cite{Minkowski:1977sc,GellMann:1980vs,Yanagida:1979as,Schechter:1980gr,Glashow:1979nm,Mohapatra:1979ia} offers a natural explanation for neutrino mass suppression. In the most straightforward seesaw realisation, heavy right-handed (RH) neutrinos are added to the SM field content (type-I seesaw), leading to an effective Majorana neutrino mass term. While these models describe qualitatively well the neutrino oscillation parameters, predicting them quantitatively becomes very difficult without further constraining the underlying model. In general, adding seesaw mediators introduces a large number of parameters at high energies, when compared to the number of low-energy observables potentially measurable by experiments. To maximize predictability, without compromising the data requirement of three nonzero mixing angles and at least two nonzero neutrino masses, two RH neutrinos must be added to the SM particle content. This is referred to as the two RH neutrino seesaw model (2RHNSM). Nevertheless, even in such minimal type-I seesaw framework, the restrictions on the high-energy Lagrangian are not sufficient to unambiguously determine the low-energy neutrino parameters.
	
It is widely established that flavour and charge-parity (CP) symmetries may strongly constrain fermion masses and mixing~\cite{Ecker:1981wv,Ecker:1987qp,Neufeld:1987wa,Grimus:1995zi,Branco:2005em,Feruglio:2012cw,Holthausen:2012dk,Girardi:2013sza,Chen:2014tpa,King:2014rwa,Ding:2014ora,Chen:2014wxa,Everett:2015oka,Branco:2015hea,Turner:2015uta,Chen:2015nha,Girardi:2015rwa,Chen:2016ica,Penedo:2017vtf,Ivanov:2017bdx,Samanta:2017kce,Chen:2018lsv}. In spite of being a demanding task~\cite{Chen:2014tpa}, imposing CP invariance in a consistent flavour framework is appealing since it may potentially lead to nontrivial constraints on both Dirac and Majorana CP-violating phases, and, ultimately, on $0\nu\beta\beta$ decay. When generalised flavour and/or CP symmetries of the theory are broken, a set of remnant CP symmetries may still be preserved by the charged-lepton and/or neutrino sectors of the Lagrangian~\cite{Chen:2014wxa,Chen:2015nha,Girardi:2015rwa}, for energies below the breaking scale. The impact of remnant CP invariance, both at high and low energies, has already been studied in the context of the 2RHNSM~\cite{Li:2017zmk}. In particular, it has been shown that the $3 \times 2$ orthogonal matrix $\OCI$ that establishes a connection between high and low energy parameters in the 2RHNSM (in the so-called Casas-Ibarra parametrisation~\cite{Casas:2001sr}) depends on a single real parameter in the presence of a remnant CP symmetry.

When texture zeros (which may be motivated by flavour symmetries~\cite{Grimus:2004hf,Dighe:2009xj,Adhikary:2009kz,Dev:2011jc,Felipe:2014vka,Cebola:2015dwa,Samanta:2015oqa,Kobayashi:2018zpq,Rahat:2018sgs,Nath:2018xih}) are considered in the Yukawa coupling and mass matrices of the 2RHNSM, the low-energy CP-violating phase space is substantially reduced~\cite{Frampton:2002qc,Ibarra:2003up,Harigaya:2012bw,Rink:2016vvl,Shimizu:2017fgu,Barreiros:2018ndn,Alcaide:2018vni}. In this context, the presence of a texture zero in the Dirac neutrino Yukawa coupling matrix fixes the single complex parameter in $\OCI$, in terms of low-energy observables~\cite{Barreiros:2018ndn}. Combining texture zeros with a remnant CP symmetry will, in principle, lead to more stringent relations among the neutrino parameters. Furthermore, predictions for the CP-violating Dirac and Majorana phases may be obtained, which could be tested in future neutrino experiments~\cite{Branco:2011zb}. Thus, a thorough study of texture zeros in the presence of remnant CP symmetries is called for. 
	
In this work, we study the 2RHNSM with one remnant CP symmetry in the neutrino sector when the maximum number of texture zeros is imposed on the lepton Yukawa and mass matrices. We analyse all possible texture zero patterns in the light of current neutrino oscillation data. The layout of the paper is as follows. In section~\ref{sec2}, we briefly review the scenario of a single remnant CP symmetry in the neutrino sector of the general type-I seesaw model. Using the Casas-Ibarra parametrisation, we analyse the restrictions stemming from imposing that symmetry and apply them to the 2RHNSM. In section~\ref{sec3}, we identify the maximally restricted texture-zero cases and study their compatibility with current neutrino data. The predictions for the low-energy CP-violating phases and the effective mass parameter relevant in $0\nu\beta\beta$ decay experiments are given in section~\ref{sec4}. Finally, in section~\ref{sec5}, we summarise our results and present the concluding remarks.

\section{Remnant CP symmetry in the type-I seesaw model}
\label{sec2}
Adding $N$ RH neutrino fields $\nu_R$ to the SM particle content, leads to the leptonic Lagrangian $\mathcal{L}=\mathcal{L}_{\ell}+\mathcal{L}_\nu+\mathcal{L}_\text{CC}$, with

\begin{align}
\mathcal{L}_\nu&=-\overline{\ell_{L}}\Ynu \tilde{\Phi}\nu_{R} -\frac{1}{2}\overline{(\nu_{R})^c}\Mr\nu_{R} + \text{H.c.}\,,	\label{LtypeI}\\
\mathcal{L}_\ell&=-\overline{\ell_{L}}\Yl \Phi\,e_{R}  + \text{H.c.}\,,\label{Lcl}\\
\mathcal{L}_\text{CC}&=\dfrac{g}{\sqrt{2}}\overline{e_{L}}\gamma^\mu \nu_{L}W_\mu^-  + \text{H.c.}\,.\label{Lcc}
\end{align}
Here, $\ell_L\equiv (\nu_L,\, e_L)^T$ is a left-handed (LH) lepton doublet, $\Phi\equiv(\phi^+,\, \phi^0)^T$ is the SM Higgs doublet with $\tilde{\Phi}=i\sigma_2\Phi^\ast$, $e_R$ are the RH charged-lepton fields, and $W^\pm_\mu$ are the charged gauge boson fields. The $3\times3$ ($3\times N$) general complex matrix $\Yl$ ($\Ynu$) is the charged-lepton (Dirac neutrino) Yukawa coupling matrix, while $\Mr$ stands for the complex symmetric $N\times N$ RH neutrino mass matrix. The effective neutrino mass matrix $\Mnu$, generated after electroweak symmetry breaking, is given by the type-I seesaw formula~\cite{Minkowski:1977sc},
\begin{align}
\Mnu=-v^2\Ynu^{}\Mr^{-1}\Ynu^T,
\label{Mnuseesaw}
\end{align}
where $v\equiv \langle \phi^0 \rangle \simeq 174$~GeV. Being symmetric, $\Mnu$ can be diagonalised by the unitary matrix $\Unu$, such that
\vspace*{-0.3cm}
\begin{align}
\Unu^T \Mnu \Unu=\text{diag}(m_1,m_2,m_3)\equiv \mathbf{d}_m\,,
\label{Mnudiag}
\end{align}
where $m_i$ are the real and positive light neutrino masses. Rotating the LH neutrino and charged-lepton fields to their physical basis by $\nu_L\rightarrow\Unu\,\nu_L$ and $e_L\rightarrow \Ul \,e_L$, a flavour misalignment arises in the charged-current interactions $\mathcal{L}_\text{CC}$. Lepton mixing is then encoded in the unitary matrix $\U$,
\vspace*{-0.5cm}
\begin{align}
\U=\Ul^\dag \Unu\,,
\label{UPMNS}
\end{align}
which can be parametrised as~\cite{Schechter:1980gr,Rodejohann:2011vc}\footnote{As shown in these references, the phase relations between $\delta$ and $\alpha_{21,31}$ with $\phi_{12}$, $\phi_{13}$ and $\phi_{23}$ in the so-called symmetrical parameterization are $\delta=\phi_{13}-\phi_{12}-\phi_{23}$, $\alpha_{21}=-\phi_{12}$ and $\alpha_{31}=-\phi_{13}$, up to unphysical phases.}
\begin{gather}
\mathbf{U}=\begin{pmatrix}
c_{12}c_{13}&s_{12}c_{13}&s_{13}\\
-s_{12}c_{23}-c_{12}s_{23}s_{13}e^{i\delta}&c_{12}c_{23}-s_{12}s_{23}s_{13}e^{i\delta}&s_{23}c_{13}e^{i\delta}\\
s_{12}s_{23}-c_{12}c_{23}s_{13}e^{i\delta}&-c_{12}s_{23}-s_{12}c_{23}s_{13}e^{i\delta}&c_{23}c_{13}e^{i\delta}
\end{pmatrix}\textbf{P}\,,
\label{Uparam}
\end{gather}
where $\theta_{ij}$ ($i<j=1,2,3$) are the lepton mixing angles (with $s_{ij}\equiv\sin\theta_{ij}$, $c_{ij}\equiv\cos\theta_{ij}$), $\delta$ is the Dirac CP-violating phase and
\vspace*{-0.3cm}
\begin{align}
\textbf{P}=\text{diag}\left(1,e^{i\alpha_{21}},e^{i\alpha_{31}}\right)\,,
\label{Pmatrix}
\end{align}
being $\alpha_{21,31}$ Majorana phases. Currently, $\theta_{ij}$, $\delta$ and the two mass-squared differences $\dmsol=m_2^2-m_1^2$ and $\dmatm=m_3^2-m_1^2$, are constrained by data as shown in table~\ref{datatable}~\cite{deSalas:2017kay} (see also refs.~\cite{Esteban:2016qun,Capozzi:2016rtj}). Since the sign of $\dmatm$ is not yet determined, the results are shown for the two possible neutrino mass orderings: 
\begin{align}
\text{Normal ordering (NO):} & \;\;m_1 < m_2 < m_3 \;\;(\dmatm >0)\; , \\
\text{Inverted ordering (IO):} & \;\;m_3 < m_1 < m_2 \;\;(\dmatm <0)\;.
\end{align}
\begin{table}[t!]
\centering
\setlength\extrarowheight{2pt}
\begin{tabular}{lK{4.2cm}K{4.2cm}}
\hline\hline
Parameter&Best Fit $\pm1\sigma$&$3\sigma$ range\\
\hline
$\theta_{12}\;(^{\circ})$ [NO] [IO]&$34.5_{-1.0}^{+1.1}$&$31.5\rightarrow38.0$\\[0.15cm]
$\theta_{23}\;(^{\circ})$ [NO] &$41.0\pm1.1$&$38.3\rightarrow52.8$\\
$\theta_{23}\;(^{\circ})$ [IO] &$50.5\pm1.0$&$38.5\rightarrow53.0$\\[0.15cm]
$\theta_{13}\;(^{\circ})$ [NO] &$8.44_{-0.15}^{+0.18}$&$7.9\rightarrow8.9$\\
$\theta_{13}\;(^{\circ})$ [IO] &$8.41_{-0.17}^{+0.16}$&$7.9\rightarrow8.9$\\[0.15cm]
$\delta\;(^{\circ})$ [NO] &$252_{-36}^{+56}$&$0\rightarrow360$\\
$\delta\;(^{\circ})$ [IO] &$259_{+47}^{+41}$&$0\rightarrow31$ \& $142\rightarrow360$ \\[0.15cm]
$\Delta m_{21}^2\;(\times 10^{-5}\;\text{eV}^2)$  [NO] [IO]&$7.56\pm0.19$&$7.05\rightarrow8.14$\\[0.15cm]
$|\Delta m_{31}^2|\;(\times 10^{-3}\;\text{eV}^2)$ [NO] &$2.55\pm0.04$&$2.43\rightarrow2.67$\\
$|\Delta m_{31}^2|\;(\times 10^{-3}\;\text{eV}^2)$ [IO] &$2.49\pm0.04$&$2.37\rightarrow2.61$\\
\hline\hline
\end{tabular}
\caption{Neutrino oscillation parameters obtained from the global analysis of ref.~\cite{deSalas:2017kay} (see also refs.~\cite{Esteban:2016qun} and \cite{Capozzi:2016rtj}).}
\label{datatable}
\end{table} 
Despite providing a very elegant and consistent way of explaining neutrino masses, the type-I seesaw model lacks in predictability. This can be easily concluded by counting the parameters in the neutrino sector. The Dirac Yukawa coupling matrix $\Ynu$ is described by $6N$ real parameters, namely, $3N$ moduli and $3N$ phases, from which three can be removed after redefining the LH charged-lepton fields. Adding the $N$ RH neutrino masses, the high-energy neutrino sector is thus described by $7N-3$ physical parameters. On the other hand, at low energies, the effective mass matrix $\Mnu$ is defined in terms of the three light neutrino masses, three mixing angles and three phases, for a total of nine physical parameters. Thus, for $N\geq2$ (required to accommodate neutrino data), the number of low-energy parameters is insufficient to uniquely determine the high-energy Lagrangian of the theory.
	
In the charged-lepton mass basis, a convenient way of parametrising $\Ynu$ is through the Casas-Ibarra parametrisation~\cite{Casas:2001sr},
\begin{align}
\mathbf{Y}^\nu=v^{-1}\U^*\,\mathbf{d}_m^{1/2}\,\OCI\,\mathbf{d}_M^{1/2} \Ur^\dagger\,,
\label{CasasandIbarra}
\end{align}
where $\mathbf{d}_M\equiv\text{diag}(M_1,\cdots,M_N)$ and $M_i$ are the real and positive heavy neutrino masses $M_i$, obtained after diagonalising $\Mr$ with a unitary matrix $\Ur$, such that
\begin{align}
\Ur^T \Mr \Ur=\mathbf{d}_M\, .
\label{mrdiagonalization}
\end{align}
Obviously, in the heavy neutrino mass basis $\Mr=\mathbf{d}_M$, $\Ur=\mathbb{1}$ and, from eq.~\eqref{CasasandIbarra}, one has $\Ynu\equiv \Ytnu=v^{-1}\U^*\,\mathbf{d}_m^{1/2}\,\OCI\,\mathbf{d}_M^{1/2}$. Also, $\OCI$ is an orthogonal $3\times N$ complex matrix parametrised by two real parameters for $N=2$, and by $6(N-2)$ real parameters for $N\geq 3$~\cite{Heeck:2012fw}. Since any matrix $\OCI$ obeying $\OCI \OCI^T=\mathbb{1}$ leads to the same low-energy $\Mnu$, determining $\OCI$ implies considering further assumptions.
	
Imposing CP invariance of the theory provides an attractive way of increasing predictability. At high energies, the lepton sector of a given model may be invariant under a generalised CP symmetry~\cite{Grimus:1995zi} which, after spontaneous symmetry breaking, may result into remnant CP symmetries of the charged-lepton and neutrino sectors. In this work, we assume that a single remnant CP symmetry of the form
\begin{align}
\quad\nu_L&\rightarrow i\, \Xnu\,\gamma_0\,\nu_L^c\; ,
\label{CPnuL}\\
\quad\nu_R&\rightarrow i\, \XN\,\gamma_0\,\nu_R^c\; ,
\label{CPnuR}
\end{align}
is preserved by the Lagrangian of eq.~\eqref{LtypeI}. Here, $\nu_{L,R}^c\equiv C\,\bar{\nu}_{L,R}^T$, being $C$ the charge-conjugation matrix. The transformation matrices $\Xnu$ and $\XN$ are $3\times3$ and $N\times N$ unitary complex matrices, respectively. In case the Lagrangian of eq.~\eqref{LtypeI} is invariant under the above symmetries, $\Ynu$ and $\Mr$ must satisfy the conditions
\begin{align}
\Xnu^\dagger\Ynu\XN&=\Ynu^\ast,
\label{CPYnu}\\
\XN^T\Mr\XN&=\Mr^\ast\,,
\label{CPMr}
\end{align}
which, together with eq.~\eqref{Mnuseesaw}, imply 
\begin{align}
\Xnu^\dagger\Mnu\Xnu^*=\Mnu^\ast.
\label{Mnuconst}
\end{align}
From eqs.~\eqref{Mnudiag} and~\eqref{Mnuconst}, one may also see that $\Unu$ obeys 
\begin{align}
\Unu^T\Xnu\Unu&=\Xhnu=\text{diag}(\pm1,\pm1,\pm1)\,,
\label{Unuconst}
\end{align}
where in the last equality we have taken into account the fact that the light neutrinos are nondegenerate (the $\pm$ signs in the above equation are independent). In such case, besides being unitary, $\Xnu$ is also symmetric~\cite{Feruglio:2012cw}. If one of the neutrinos is massless, $\Xhnu$ has the more general form
\begin{align}
	\Xhnu=\begin{cases}
		\text{diag}(e^{i\phi},a,b)\;\;\text{for NO}\\
		\text{diag}(a,b, e^{i\phi})\;\;\text{for IO}
	\end{cases}\!\!\!\!\!\!, \,\,(a,b)=(\pm 1,\pm 1)\,,
\label{XnuIONO}
\end{align}
where the presence of the phase factor $e^{i\phi}$, instead of $\pm 1$, is due to $m_1=0$ ($m_3=0$) for NO (IO). For a given $\Unu$, eq.~\eqref{Unuconst} fixes the CP transformation matrix $\Xnu$ as $\Xnu=\Unu^\ast \Xhnu \Unu^\dag$ which, in the charged-lepton mass basis takes the form $\Xnu=\U^\ast \Xhnu \U^\dag$. 

For RH neutrinos, using eqs.~\eqref{mrdiagonalization} and~\eqref{CPMr}, we have
\begin{align}
\Ur^\dagger\XN\Ur^*&=\XhN\,,
\label{Urconst}
\end{align}
where $\XhN$ is automatically unitary, due to the unitarity of $\Ur$ and $\XN$. In contrast to what is observed for light neutrinos, degeneracies in the heavy Majorana neutrino mass spectrum must be taken into account. For $n \leq N$ degenerate RH neutrinos, i.e. $M_1=M_2=...=M_n$, the solution of the above equation is
\begin{align}
	\XhN=\begin{pmatrix}
	\mathbf{O}_n & 0\\
		0        &\mathbf{D}_{N-n}
    \end{pmatrix}\,,
	\label{XNhatdef}
\end{align}
where $\mathbf{O}_n$ is a $n\times n$ general real orthogonal matrix, and $\mathbf{D}_{N-n}$ is a $(N-n)\times (N-n)$ diagonal matrix with entries $\pm 1$, i.e. $\mathbf{D}_{N-n}=\text{diag}(\pm1,\pm1,\cdots,\pm1)$. In general, $\XN$ is symmetric only if $n=0$ (nondegenerate heavy neutrinos). Altogether, eqs.~\eqref{CasasandIbarra}, \eqref{CPYnu}, \eqref{CPMr}, \eqref{Unuconst} and \eqref{Urconst} lead to the following constraining relation for $\OCI$:
\begin{align}
\OCI=\Xhnu\OCI^*\XhN^\dagger\; .
\label{CPR}
\end{align}
In what follows, we will consider both degenerate and nondegenerate heavy neutrino mass spectra.

For nondegenerate heavy neutrinos, it can be easily shown that the remnant CP symmetry enforces the matrix elements of $\OCI$ to obey the relation $\OCI_{ij}=\pm\OCI_{ij}^*$, i.e., $\OCI_{ij}$ are either real or purely imaginary. Therefore, in this case, the number of free parameters in $\OCI$ is decreased by half, being now equal to one if $N=2$, and $3(N-2)$ if $N\geq3$. On the other hand, when all heavy neutrinos are degenerate, i.e. $n=N$ in eq.~\eqref{XNhatdef}, one has from eq.~\eqref{CPR}
\begin{align} 
\XhN \equiv \mathbf{O}_N=\OCI^T\Xhnu\OCI^\ast.
\label{ONconst}
\end{align}
This means that the matrix $\OCI$ in the Casas-Ibarra parametrisation must be such that $\mathbf{O}_N$ is a real matrix, since $\XhN$ is unitary. Obviously, if $\Xhnu=\pm \text{diag}(1,1,1)$, $\OCI$ is constrained to be real, and $\mathbf{O}_N=\pm \mathbb{1}_{N}$. In this case $\XN=\pm \mathbb{1}_{N}$ and, as expected, the CP transformation in the RH neutrino sector is trivial.
	
\subsection{The two RH neutrino case}
\label{sec2.1} 
We now consider the 2RHNSM, i.e., the minimal type-I seesaw scenario compatible with experimental data. In this case, two heavy Majorana neutrinos, with masses $M_1$ and $M_2$, are responsible for the neutrino mass generation, being one of the light neutrinos massless ($m_1=0$ for NO, or $m_3=0$ for IO). Moreover, one of the Majorana phases in eq.~\eqref{Pmatrix} can be removed by redefinition of the neutrino fields, leaving a single physical CP-violating Majorana phase $\alpha$. In our notation, this corresponds to replacing \textbf{P} in eq.~\eqref{Uparam} by 
\begin{align}
\textbf{P}=\text{diag}(1,e^{i\alpha},1)\,.
\end{align}
Hence, the number of low-energy neutrino parameters in the 2RHNSM is reduced to seven: three mixing angles, two nonzero light neutrino masses, a Dirac phase and a single Majorana phase. On the other hand, at high energies, $\Ynu$ is a $3\times2$  matrix defined by nine independent parameters, while $\Mr$ is a $2\times2$ matrix with two parameters (in the RH neutrino mass basis). Furthermore, the matrix $\OCI$ in eq.~\eqref{CasasandIbarra} is an orthogonal $3\times 2$ complex matrix parametrised by a complex angle $\hat{z}$ in the following way~\cite{Ibarra:2003up}
\begin{gather}
	\text{NO}:\;\OCI=\begin{pmatrix}
		0&0\\
		\cos \hat{z}&-\sin \hat{z}\\
		\xi \sin \hat{z}&\xi \cos \hat{z} 
	\end{pmatrix}
	\,,\qquad
	\text{IO}:\; \OCI=\begin{pmatrix}
		\cos \hat{z} &-\sin \hat{z} \\
		\xi \sin \hat{z} &\xi \cos \hat{z}\\
		0&0
	\end{pmatrix}\; ,
	\label{RmatrixIO}
\end{gather}
with $\xi=\pm 1$ resulting from a discrete indeterminacy in $\OCI$. As previously seen, the presence of CP symmetries may decrease the number of free parameters in a generic seesaw model. In the specific case of the 2RHNSM, $\Xhnu$ and $\XhN$ are given by eqs.~\eqref{XnuIONO} and \eqref{XNhatdef}, respectively, where $N=2$ must be taken.

For nondegenerate heavy neutrinos ($M_1\neq M_2$), imposing one remnant CP symmetry decreases the number of free parameters in $\Ynu$ by one (see discussion at the end of the previous subsection). Therefore, the total number of high-energy parameters is reduced to ten, which is still large when compared to the seven low-energy observables. According to eq.~\eqref{XNhatdef}, for $M_1\neq M_2$, one has  $\XhN=\text{diag}(\pm1,\pm1)$.  Since the transformations $\Xhnu\rightarrow-\Xhnu$ and $\XhN\rightarrow-\XhN$ leave eq.~\eqref{CPR} invariant, we will only consider the cases $\XhN=\text{diag}(1,\pm 1)$ in our analysis. All the allowed configurations for $\OCI$, obeying eq.~\eqref{CPR} with different forms of $\Xhnu$ and $\XhN$, are presented in table~\ref{Rforms} (see also ref.~\cite{Li:2017zmk}). As expected, for $M_1\neq M_2$, the elements $\OCI_{ij}$ are either real or purely imaginary, and only a single real parameter $z$ is required to define $\OCI$. 
\begin{table}[t!]
\centering
\setlength\extrarowheight{4pt}
\begin{tabular}{ccccc}
\hline\hline
$\XhN$&$(a,b)$&$\OCI$ (NO)&$\OCI$ (IO)&Label\\
\hline\hline\\[-0.55cm]
\multirow{4}{*}{diag$(1,1)$}&$(1,1)$&$\begin{pmatrix}
0&0\\
\cos z & -\sin z\\
\xi \sin z&\xi\cos z
\end{pmatrix}$&$\begin{pmatrix}
\cos z & -\sin z\\
\xi \sin z&\xi\cos z\\
0&0
\end{pmatrix}$&$\OI$\\
&$(1,-1)$&\xmark&\xmark&\xmark\\
&$(-1,1)$&\xmark&\xmark&\xmark\\
&$(-1,-1)$&\xmark&\xmark&\xmark\\
\hline
\multirow{9}{*}{diag$(1,-1)$}&$(1,1)$&\xmark&\xmark&\xmark\\
&$(1,-1)$&$\pm\begin{pmatrix}
0&0\\
\cosh z & -i\sinh z\\
i\xi \sinh z&\xi\cosh z
\end{pmatrix}$&$\pm\begin{pmatrix}
\cosh z & -i\sinh z\\
i\xi \sinh z&\xi\cosh z\\
0&0
\end{pmatrix}$&$\OII$\\
&$(-1,1)$&$\pm\begin{pmatrix}
0&0\\
i\sinh z & -\cosh z\\
\xi \cosh z&i\xi\sinh z
\end{pmatrix}$&$\pm\begin{pmatrix}
i\sinh z & -\cosh z\\
\xi \cosh z&i\xi\sinh z\\
0&0
\end{pmatrix}$&$\OIII$\\
&$(-1,-1)$&\xmark&\xmark&\xmark\\		
\hline\hline
\end{tabular}
\caption{Possible parametrisations for $\OCI$ in the nondegenerate case $M_1 \neq M_2$. The notation $\Xhnu=\text{diag}(e^{i\phi},a,b)$ for NO and $\Xhnu=\text{diag}(a,b, e^{i\phi})$ for IO is used. The symbol `\xmark' indicates that there is no matrix $\OCI$ fulfilling eq.~\eqref{CPR} for the corresponding $\XhN$ and $\Xhnu$.}
\label{Rforms}
\end{table}
For the degenerate case $M_1 = M_2$, one has $\mathbf{O}_N \equiv \mathbf{O}_2$ in eq.~\eqref{ONconst}, with
\begin{align}
	\mathbf{O}_2=\OCI^T\Xhnu\OCI^\ast.
	\label{ONconst2}
\end{align}
Since $\mathbf{O}_2$ must be real, the above relation restricts $\OCI$ to be of the type $\OI$ given in table~\ref{Rforms} for $(a,b)=\pm(1,1)$ in $\Xhnu$ (for notation, see the table caption or eq.~\eqref{XnuIONO}). On the other hand, if $(a,b)=\pm(1,-1)$, $\mathbf{O}_2$ is automatically real for any matrix $\OCI$.  In this case, no constraints are imposed by the CP symmetry and the number of parameters in $\Ynu$ remains unchanged.

In summary, except for the case discussed above with $M_1=M_2$ and $(a,b)=\pm(1,-1)$, the remnant CP symmetry forces the elements of $\mathbf{O}$ to be either real or purely imaginary. Thus, $\mathbf{O}$ is parametrised by a single real parameter $z$ (instead of two in the general 2RHNSM), and $\Ynu$ is described by eight independent parameters (instead of nine in the general 2RHNSM). In order to increase the degree of predictability of the 2RHNSM combined with a CP symmetry, in the following we will consider maximally restricted texture zeros in $\Yl$, $\Ynu$ and $\Mr$.
 
\section{Texture zeros in the 2RHNSM with a remnant CP symmetry}
\label{sec3}
In a previous work~\cite{Barreiros:2018ndn}, we studied maximally restricted texture zero patterns for $\Yl$, $\Ynu$ and $\Mr$ in the 2RHNSM. We concluded that the maximal number of texture zeros that can accommodate neutrino and charged-lepton data is six in $\Yl$, two in $\Ynu$ and a single zero in $\Mr$. In this scenario, only an IO light-neutrino mass spectrum turns out to be compatible with data. For each of the viable patterns, it was shown that the low-energy Dirac and Majorana phases can be expressed in terms of the well-measured mixing angles and neutrino mass-squared differences, being the results independent of the mass ratio $r_N=M_2/M_1$. We now intend to analyse the compatibility of those textures when a remnant CP symmetry is imposed. 

In the basis where $\Yl$ is diagonal, the following textures for $\Mr$ will be considered:
\begin{align}
{\rm R}_1:\begin{pmatrix}
\times&0\\
\cdot&\times
\end{pmatrix},\;
{\rm R_2}:\begin{pmatrix}
0&\times\\
\cdot&\times
\end{pmatrix},\;
{\rm R_3}:\begin{pmatrix}
\times&\times\\
\cdot&0
\end{pmatrix},\;
{\rm R_4}:\begin{pmatrix}
0&\times\\
\cdot&0
\end{pmatrix},
\label{Rstructures}
\end{align}
where `$\times$' denotes a generic nonvanishing entry and `$\cdot$' indicates the symmetric nature of the matrix. Notice that R$_4$ automatically leads to $M_1=M_2$, being the only nondiagonal pattern allowing for a mass degeneracy of the two heavy neutrinos. Therefore, only R$_1$ and R$_4$ may reproduce a degenerate RH neutrino mass spectrum. 

Let us first discuss the possibility of having two texture zeros in $\Ynu$ together with the remnant CP symmetry. The possible $\Ynu$ patterns in this case are~\cite{Barreiros:2018ndn}
\begin{align}
	\text{T}_1:\! \begin{pmatrix}
		0&\times\\
		\times&0\\
		\times&\times
	\end{pmatrix}\!,\text{T}_2:\! \begin{pmatrix}
		0&\times\\
		\times&\times\\
		\times&0
	\end{pmatrix}\!,\text{T}_3:\! \begin{pmatrix}
	\times&\times\\
	0&\times\\
	\times&0
	\end{pmatrix}\!,\text{T}_4:\! \begin{pmatrix}
		\times&0\\
		0&\times\\
		\times&\times
	\end{pmatrix}\!,\text{T}_5:\! \begin{pmatrix}
		\times&0\\
		\times&\times\\
		0&\times
	\end{pmatrix}\!,\text{T}_6:\! \begin{pmatrix}
	\times&\times\\
	\times&0\\
	0&\times
	\end{pmatrix}\!.
	\label{ynupatterns2}
\end{align}
Taking into account the results obtained in ref.~\cite{Barreiros:2018ndn}, the following conclusions hold in the present work:
\begin{itemize}
\item {For $\Mr$ of the type R$_4$, it was shown that none of the above T textures are compatible with neutrino data. This is true regardless of the presence of the CP symmetry.}
\item {For R$_1$ with $M_1=M_2$ and $(a,b)=\pm(1,-1)$ in eq.~\eqref{XnuIONO}, $\OCI$ is not constrained by the CP symmetry and the results from ref.~\cite{Barreiros:2018ndn} hold. Namely, for T$_1$, T$_2$, T$_4$ and T$_5$, which are viable for the IO case, $\delta$ can be close to the current best-fit value $\delta \simeq 3\pi/2$ with $\alpha \simeq 0.94\pi\,(0.04\pi)$ for T$_1$ and T$_4$ (T$_2$ and T$_5$).}
\item {For R$_1$ with $M_1=M_2$ and $(a,b)=\pm(1,1)$, and R$_{1,2,3}$ with $M_1 \neq M_2$, the CP symmetry constrains $\OCI$ to be parametrised by a single real parameter. As it will be shown later, some of these cases with R$_{2,3}$ are compatible with data, while those with R$_1$ are excluded.}
\end{itemize}

In view of the above, and for those cases in which two texture zeros are excluded, we will now turn our attention to the one-texture zero $\Ynu$ patterns 
\begin{align}
\text{Y}_1:\! \begin{pmatrix}
0&\times\\
\times&\times\\
\times&\times
\end{pmatrix}\!,\text{Y}_2:\! \begin{pmatrix}
\times&\times\\
0&\times\\
\times&\times
\end{pmatrix}\!,\text{Y}_3:\! \begin{pmatrix}
\times&\times\\
\times&\times\\
0&\times
\end{pmatrix}\!,\text{Y}_4:\! \begin{pmatrix}
\times&0\\
\times&\times\\
\times&\times
\end{pmatrix}\!,\text{Y}_5:\! \begin{pmatrix}
\times&\times\\
\times&0\\
\times&\times
\end{pmatrix}\!,\text{Y}_6:\! \begin{pmatrix}
\times&\times\\
\times&\times\\
\times&0
\end{pmatrix}\!.
\label{ynupatterns}
\end{align}
%
%\begin{table}
%	\centering
%	\setlength\extrarowheight{3pt}
%	\begin{tabular}{ccK{1cm}K{1cm}}
%		\hline\hline
%		\multirow{2}{*}{$\Ynu$}&\multirow{2}{*}{$\Mnu$}&\multicolumn{2}{c}{Compatibility}\\
%		&&NO&IO\\
%		\hline\\[-0.45cm]
%		Y$_{1,4}$&A: $\begin{pmatrix}
%		0 &\times&\times\\
%		\cdot&\times&\times\\
%		\cdot&\cdot&\times
%		\end{pmatrix}$&\xmark&\xmark\tspp\bspp\\\\[-0.45cm]
%		Y$_{2,5}$&D: $\begin{pmatrix}
%		\times &\times&\times\\
%		\cdot&0&\times\\
%		\cdot&\cdot&\times
%		\end{pmatrix}$&\xmark&\cmark (3$\sigma$)\tspp\bspp\\\\[-0.45cm]
%		Y$_{3,6}$&F: $\begin{pmatrix}
%		\times &\times&\times\\
%		\cdot&\times&\times\\
%		\cdot&\cdot&0
%		\end{pmatrix}$&\xmark&\cmark (3$\sigma$)\tspp\bspp\\\\[-0.5cm]
%		\hline\hline
%	\end{tabular}
%	\caption{Textures for the effective neutrino mass matrix $\Mnu$ obtained with the seesaw formula~\eqref{Mnuseesaw}, considering the $\Ynu$ textures of eq.~\eqref{ynupatterns} and the pattern R$_4$ for $\Mr$. The `\cmark' and `\xmark' symbols indicate whether or not the corresponding texture is compatible with data.}
%	\label{tabR4}
%\end{table}
%
\begin{table}
	\centering
	\setlength\extrarowheight{3pt}
	\begin{tabular}{ccK{1cm}K{1cm}K{8cm}}
		\hline\hline
		\multirow{2}{*}{$\Ynu$}&\multirow{2}{*}{$\Mnu$}&\multicolumn{2}{c}{Compatibility}&Predictions\\
		&&NO&IO&IO (compatible cases)\\
		\hline\\[-0.45cm]
		Y$_{1,4}$&A: $\begin{pmatrix}
		0 &\times&\times\\
		\cdot&\times&\times\\
		\cdot&\cdot&\times
		\end{pmatrix}$&\xmark&\xmark&$-$\tspp\bspp\\\\[-0.45cm]
		Y$_{2,5}$&D: $\begin{pmatrix}
		\times &\times&\times\\
		\cdot&0&\times\\
		\cdot&\cdot&\times
		\end{pmatrix}$&\xmark&\cmark (3$\sigma$)& $\cos\delta=2\frac{(c^2_{12}\sqrt{1+r_\nu}-s^2_{12})c^2_{23}+(s^2_{12}\sqrt{1+r_\nu}-c^2_{12})s^2_{23}s^2_{13}}{(\sqrt{1+r_\nu}+1)\sin(2\theta_{12})\sin(2\theta_{23})s_{13}}$ $\cos(2\alpha)\simeq-\frac{3+\cos(4\theta_{12})-16s_{13}^2t_{23}^2}{2\sin^2(2\theta_{12})}$\tspp\bspp\\\\[-0.45cm]
		Y$_{3,6}$&F: $\begin{pmatrix}
		\times &\times&\times\\
		\cdot&\times&\times\\
		\cdot&\cdot&0
		\end{pmatrix}$&\xmark&\cmark (3$\sigma$)& $\cos\delta=2\frac{(s^2_{12}-c^2_{12}\sqrt{1+r_\nu})s^2_{23}+(c^2_{12}-s^2_{12}\sqrt{1+r_\nu})c^2_{23}s^2_{13}}{(\sqrt{1+r_\nu}+1)\sin(2\theta_{12})\sin(2\theta_{23})s_{13}}$ $\cos(2\alpha)\simeq-\frac{3t_{23}^2+t_{23}^2\cos(4\theta_{12})+16s_{13}^2}{2t_{23}^2\sin^2(2\theta_{12})}$\tspp\bspp\\\\[-0.5cm]
		\hline\hline
	\end{tabular}
	\caption{Textures for the effective neutrino mass matrix $\Mnu$ obtained with the seesaw formula~\eqref{Mnuseesaw}, considering the $\Ynu$ textures of eq.~\eqref{ynupatterns} and the pattern R$_4$ for $\Mr$. The `\cmark' and `\xmark' symbols indicate whether or not the corresponding texture is compatible with data. Hereafter, $r_\nu \equiv \dmsol/|\dmatm|$.}
	\label{tabR4}
\end{table}
Combining $\Mr$ of the type R$_4$ ($M_1=M_2$) with the above textures using eq.~\eqref{Mnuseesaw}, we obtain the $\Mnu$ patterns (A, D and F in the notation of ref.~\cite{Barreiros:2018ndn}) shown in the second column of table~\ref{tabR4}. Notice that, in all cases, $\Mnu$ features one texture zero in a diagonal entry, which lead to testable relations among the low-energy parameters (see eq.~\eqref{Mnudiag}). As concluded in ref.~\cite{Barreiros:2018ndn} and shown in the third and fourth columns of table~\ref{tabR4}, texture D (F) is compatible with neutrino data at $3\sigma$  for an IO neutrino mass spectrum\footnote{A given texture combination is considered compatible with data at $3\sigma$ when all predicted parameters lie in the $3\sigma$ range given in table~\ref{datatable} and at least one is out of the $1\sigma$ interval.} with $\delta$ varying in the range $[-0.3,0.3]\pi$ ($[0.7,1.3]\pi$) and $\alpha$ in the ranges $[0.4,0.6]\pi$ and $[1.4,1.6]\pi$, according to the predictions in the fifth column of table~\ref{tabR4}. These results remain valid in the present framework as long as the remnant CP symmetry does not introduce further constraints on the model parameters, which is the case for $(a,b)=\pm(1,-1)$ (unconstrained $\OCI$). In conclusion, only textures Y$_2$, Y$_5$, Y$_3$ and Y$_6$ are viable when $\Mr$ is of the R$_4$ type and $(a,b)=\pm(1,-1)$. The remaining texture combinations with $M_1=M_2$ ($\Mr$ of type R$_{1,4}$) and $(a,b)=\pm(1,1)$ will be discussed later.

Let us now study the case $M_2\neq M_1$, for which only patterns R$_1$, R$_2$ and R$_3$ need to be considered. As we shall see later, the most restrictive case of two texture zeros in $\Ynu$ and the existence of a remnant CP symmetry in the 2RHNSM is compatible with current neutrino data only for $\Mr$ of the types R$_{2,3}$. To understand this more easily, we start by analysing the patterns of $\Ynu$ with one texture zero. 

For the specific combinations $(\text{R}_{2},\text{Y}_{1-3})$ and $(\text{R}_{3},\text{Y}_{4-6})$, a single texture zero arises in one of the diagonal entries of $\Mnu$. This is true for any complex $\hat{z}$ parametrising $\OCI$ in eq.~\eqref{CasasandIbarra}. If $\Mr$ is of the type R$_2$ (R$_3$) and the $\Ynu$ pattern corresponds to Y$_1$, Y$_2$ or Y$_3$ (Y$_4$, Y$_5$ or Y$_6$), textures A, D or F will arise for $\Mnu$, respectively, according to the notation in table~\ref{tabR4}. As discussed above, only textures D and F are compatible with data at $3\sigma$, with $\alpha$ and $\delta$ constrained to obey the relations in the fifth column table~\ref{tabR4}. Instead, for all the remaining combinations $(\text{R}_{1},\text{Y}_{1-6})$, $(\text{R}_{2},\text{Y}_{4-6})$ and $(\text{R}_{3},\text{Y}_{1-3})$, no zeros (or any other constraint) arise in $\Mnu$, so that the presence of a single texture zero in $\Ynu$ does not restrict the low-energy parameter space. Nevertheless, for these cases, imposing a texture zero in $\Ynu$ leads to the determination of $\hat{z}$ in eq.~\eqref{RmatrixIO} in terms of the low-energy neutrino parameters~\cite{Barreiros:2018ndn} (and $M_{1,2}$ for a non-diagonal $\Mr$), fixing the matrix $\OCI$ in eq.~\eqref{CasasandIbarra}.  
For instance, in the simplest case of diagonal $\Yl$ and $\Mr$, the constraint $(\Ytnu)_{11}=0$ together with eq.~\eqref{CasasandIbarra} and $\OCI$ given by eq.~\eqref{RmatrixIO}, leads to
\begin{align}
\text{NO:}&\quad\sqrt{m_2}\,\U_{12}^*c_{\hat{z}}+\xi\sqrt{m_3}\,\U_{13}^*s_{\hat{z}}=0\; ,\\
\text{IO:}&\quad\sqrt{m_1}\,\U_{11}^*c_{\hat{z}}+\xi\sqrt{m_2}\,\U_{12}^*s_{\hat{z}}=0\; ,
\end{align}
where $c_{\hat{z}}\equiv \cos\hat{z}$ and $s_{\hat{z}}\equiv \sin\hat{z}$. The above relations determine $\hat{z}$ as
\begin{align}
\text{NO:}&\quad\tan \hat{z}=-\xi\sqrt{\dfrac{m_2}{m_3}}\dfrac{\U_{12}^*}{\U_{13}^*}\; ,\\
\text{IO:}&\quad\tan \hat{z}=-\xi\sqrt{\dfrac{m_1}{m_2}}\dfrac{\U_{11}^*}{\U_{12}^*}\; .
\end{align}
In any of the cases  $(\text{R}_{1-3},\text{Y}_{1-6})$, the number of independent parameters in $\Ynu$ is reduced from nine to seven, due to the presence of two low-energy constraints, or due to the determination of the complex parameter $\hat{z}$.

Although a single texture zero in $\Ynu$ may not lead to low-energy constraints in the general 2RHNSM, this is not the case when a remnant CP symmetry is considered. In this framework (see section~\ref{sec2.1}), and according to eq.~\eqref{CPR}, the matrix $\OCI$ is parametrised by a real parameter $z$, reducing the number of parameters in $\Ynu$ from seven to six. Hence, we are left with eight high-energy parameters to be compared with seven low-energy observables. As shall be seen below, depending on the constraints imposed on $\Ynu$ and $\Mr$, some of the physical parameters can be related to each other and, therefore, low-energy predictions are obtained.

We now turn to the compatibility analysis of the patterns given in eq.~\eqref{ynupatterns} when $\Mr$ exhibits the forms R$_{1,2,3}$. Depending on the column in which the zero in $\Ynu$ is placed, the following conditions are obtained
\begin{align}
(\Ynu)_{j 1}=0:&\quad (\Ytnu)_{j 1}\left(\Ur^*\right)_{11}+(\Ytnu)_{j 2}(\Ur^*)_{12}=0\, ,
\label{Ynua1}\\
(\Ynu)_{j 2}=0:&\quad (\Ytnu)_{j 1}(\Ur^*)_{21}+(\Ytnu)_{j 2}(\Ur^*)_{22}=0\,,
\label{Ynua2}
\end{align}
where $\Ur$, defined in eq.~\eqref{mrdiagonalization}, depends solely on the ratio $r_N=M_2/M_1$, and $\Ytnu$ corresponds to $\Ynu$ defined in the heavy neutrino mass basis. Using the parametrisation (\ref{CasasandIbarra}), one has
\begin{align}
(\Ytnu)_{j 1}=\begin{cases}
\dfrac{\sqrt{M_1}}{v}\left(\sqrt{m_2}\,\U_{j 2}^*\OCI_{21}+\sqrt{m_3}\,\U_{j 3}^*\OCI_{31}\right)\,,\quad \text{NO}\,,\\[0.3cm]
\dfrac{\sqrt{M_1}}{v}\left(\sqrt{m_1}\,\U_{j 1}^*\OCI_{11}+\sqrt{m_2}\,\U_{j 2}^*\OCI_{21}\right)\,,\quad \text{IO}\,,
\end{cases}\\
(\Ytnu)_{j 2}=\begin{cases}
\dfrac{\sqrt{M_2}}{v}\left(\sqrt{m_2\,}\U_{j 2}^*\OCI_{22}+\sqrt{m_3}\,\U_{j 3}^*\OCI_{32}\right)\,,\quad \text{NO}\,,\\[0.3cm]
\dfrac{\sqrt{M_2}}{v}\left(\sqrt{m_1}\,\U_{j 1}^*\OCI_{12}+\sqrt{m_2}\,\U_{j 2}^*\OCI_{22}\right)\,,\quad \text{IO}\,.
\end{cases}
\end{align}
When the remnant CP symmetry is imposed, $\OCI$ is constrained to be of the forms $\mathbf{O}_{\text{I-III}}$ (see table~\ref{Rforms}). Replacing in each case the matrix elements of $\OCI$ in eqs.~\eqref{Ynua1} and~\eqref{Ynua2}, an expression for $\tan z$ (for $\OI$) or $\tanh z$ (for $\OII$ and $\OIII$) is obtained. Since $z$ has to be real, those expressions must fulfill the conditions
\begin{align}
\OI: &\;\text{Im}[\tan z]=0 \, ,
\label{condR1st}\\
\OII,\OIII: &\;\text{Im}[\tanh z]=0\,,\,|\tanh z|<1\, .
\label{condR2ndR3rd}
\end{align}
\begin{table}[t!]
	\centering
	\setlength\extrarowheight{1pt}
	\begin{tabular}{K{1.8cm}K{0.3cm}K{11.7cm}}
		\hline\hline
		($\Mr$,$\Ynu$)&&Constraints for $\OCI=\OI$ ($\text{Im}[\tan z]=0$)\\
		\hline\hline	
		\multirow{3}{*}[0.1cm]{(R$_1$,Y$_{1,4}$)}&NO&$\sin\alpha=0$\bsp\tsp\\
		&IO&$\sin\alpha=0$\bspp\tspp\\
		\hline
		\multirow{3}{*}[0.1cm]{(R$_1$,Y$_{2,5}$)}&NO&$\tan\alpha=-\frac{\sin\delta }{t_{12}t_{23}s_{13}- \cos\delta}$\bsp\tsp\\
		&IO&$\tan\alpha=\frac{4\sin(2\theta_{23}) s_{13}\sin\delta}{\sin(2\theta_{12})[1+3\cos(2\theta_{23})+2\cos(2\theta_{13})s_{23}^2]+4\cos(2\theta_{12})\sin(2\theta_{23})s_{13}\cos\delta}$\bsp\tsp\\	
		\hline
		\multirow{3}{*}[0.1cm]{(R$_1$,Y$_{3,6}$)}&NO&$\tan\alpha=\frac{t_{23}\sin\delta }{t_{12}s_{13}+ t_{23}\cos\delta}$\bsp\tsp\\
		&IO&$\tan\alpha=-\frac{4\sin(2\theta_{23}) s_{13}\sin\delta}{\sin(2\theta_{12})[1-3\cos(2\theta_{23})+2\cos(2\theta_{13})c_{23}^2]-4\cos(2\theta_{12})\sin(2\theta_{23})s_{13}\cos\delta}$\bsp\tsp\\
		\hline
		\multirow{3}{*}[0.1cm]{\specialcell{(R$_2$,Y$_4$)\\(R$_3$,Y$_1$)}}&NO&$\sin\alpha=\frac{r_N}{r_N^2+1}\frac{s_{13}^2+\sqrt{r_\nu}\,c_{13}^2s_{12}^2}{\xi\sqrt[4]{r_\nu}\, c_{13}s_{13}s_{12}}$\bsp\tsp\\
		&IO&$\sin\alpha=-\frac{r_N}{r_N^2+1}\frac{(\sqrt{1+r_\nu}+1)-(\sqrt{1+r_\nu}-1)\cos(2\theta_{12})}{\xi\sqrt[4]{1+r_\nu} \sin(2\theta_{12})}$\bsp\tsp\\
		\hline
		\multirow{3}{*}[0.1cm]{\specialcell{(R$_2$,Y$_5$)\\(R$_3$,Y$_2$)}}&NO&$\sin\left(\alpha-\delta\right)\simeq\frac{r_N}{r_N^2+1}\frac{2(\sqrt{r_\nu}\,c_{23}^2c_{12}^2+s_{23}^2)}{\xi\sqrt[4]{r_\nu}\, c_{12}\sin(2\theta_{23})}$\bsp\tsp\\
		&IO&$\sin\alpha\simeq\frac{r_N}{r_N^2+1}\frac{2(\sqrt{1+r_\nu}\,c_{12}^2+s_{12}^2)}{\xi\sqrt[4]{1+r_\nu} \sin(2\theta_{12})}$\bsp\tsp\\	
		\hline
		\multirow{3}{*}[0.1cm]{\specialcell{(R$_2$,Y$_6$)\\(R$_3$,Y$_3$)}}&NO&$\sin\left(\alpha-\delta\right)\simeq-\frac{r_N}{r_N^2+1}\frac{2(\sqrt{r_\nu}\,s_{23}^2c_{12}^2+c_{23}^2)}{\xi\sqrt[4]{r_\nu}\, c_{12}\sin(2\theta_{23})}$\bsp\tsp\\
		&IO&$\sin\alpha\simeq\frac{r_N}{r_N^2+1}\frac{2(\sqrt{1+r_\nu}\,c_{12}^2+s_{12}^2)}{\xi\sqrt[4]{1+r_\nu} \sin(2\theta_{12})}$\bsp\tsp\\				
		\hline\hline
	\end{tabular}
	\caption{Relations among low-energy parameters obtained in the 2RHNSM with one remnant CP symmetry, for $\Ynu$ and $\Mr$ of the forms Y$_{1-6}$ and R$_{1-3}$, respectively. Here, $r_N\neq 1$ and $\OCI=\OI$. Notice that, since interchanging the second and third lines of $\Ynu$ corresponds to the same operation in $\U$ (see eq.~\eqref{CasasandIbarra}), the results for Y$_{3}$ (Y$_{6}$) are obtained from those of Y$_{2}$ (Y$_{5}$) performing the transformation $\theta_{23}\rightarrow \theta_{23}+\pi/2$. This can be seen by comparing the second (sixth) and third (seventh) rows of the table. These properties are general and, therefore, are also verified in table~\ref{condIm0R2ndR3rd}. Due to the complexity of the relations for $\OI$, only the zeroth order in $s_{13}$ is shown for the results of R$_{2,3}$ with the patterns Y$_{2,3,5,6}$. }
	\label{condIm0R1st}
\end{table}
\begin{table}[t!]
	\centering
	\setlength\extrarowheight{1pt}
	\begin{tabular}{K{1.8cm}K{0.3cm}K{11.7cm}}
		\hline\hline
		($\Mr$,$\Ynu$)&&Constraints for $\OCI=\OII,\OIII$ ($\text{Im}[\tanh z]=0$)\\
		\hline\hline		
		\multirow{2}{*}{\specialcell{(R$_1$,Y$_{1,4}$)\\(R$_2$,Y$_4$)\\(R$_3$,Y$_1$)}}&NO&$\cos\alpha=0$\bspp\tspp\\
		&IO&$\cos\alpha=0$\bspp\tspp\\
		\hline
		\multirow{2}{*}{\specialcell{(R$_1$,Y$_{2,5}$)\\(R$_2$,Y$_5$)\\(R$_3$,Y$_2$)}}&NO&$\tan\alpha=\frac{t_{12}t_{23}s_{13}-\cos\delta}{\sin\delta }$\bspp\tspp\\
		&IO&$\tan\alpha=-\frac{\sin(2\theta_{12})[1+3\cos(2\theta_{23})+2\cos(2\theta_{13})s_{23}^2]+4\cos(2\theta_{12})\sin(2\theta_{23})s_{13}\cos\delta}{4\sin(2\theta_{23}) s_{13}\sin\delta}$\bspp\tspp\\	
		\hline
		\multirow{2}{*}{\specialcell{(R$_1$,Y$_{3,6}$)\\(R$_2$,Y$_6$)\\(R$_3$,Y$_3$)}}&NO&$\tan\alpha=-\frac{t_{12}s_{13}+t_{23}\cos\delta}{t_{23}\sin\delta }$\bspp\tspp\\
		&IO&$\tan\alpha=\frac{\sin(2\theta_{12})[1-3\cos(2\theta_{23})+2\cos(2\theta_{13})c_{23}^2]-4\cos(2\theta_{12})\sin(2\theta_{23})s_{13}\cos\delta}{4\sin(2\theta_{23}) s_{13}\sin\delta}$\bspp\tspp\\								
		\hline\hline		
	\end{tabular}
	\caption{As in table~\ref{condIm0R1st}, for the cases $\OCI=\OII,\OIII$. }
	\label{condIm0R2ndR3rd}
\end{table}

For the combinations $(\text{R}_{2},\text{Y}_{1-3})$ and $(\text{R}_{3},\text{Y}_{4-6})$, the texture zero in $\Ynu$ is guaranteed if the low-energy relations $(\Mnu)_{11}=0$, $(\Mnu)_{22}=0$ or $(\Mnu)_{33}=0$ hold, regardless of $z$. Thus, as long as the conditions in eqs.~\eqref{condR1st} or~\eqref{condR2ndR3rd} are fulfilled, the CP symmetry is ensured. As concluded in ref.~\cite{Barreiros:2018ndn}, only the pattern combinations which lead to $(\Mnu)_{22}=0$ and $(\Mnu)_{33}=0$, respectively, are viable at 3$\sigma$ for IO (see table~\ref{tabR4}). In contrast, for all remaining cases $(\text{R}_{1},\text{Y}_{1-6})$, $(\text{R}_{2},\text{Y}_{4-6})$ and $(\text{R}_{3},\text{Y}_{1-3})$, the texture zero in $\Ynu$ fixes $z$ in terms of low-energy neutrino parameters (and $r_N$) through eqs.~\eqref{Ynua1} or~\eqref{Ynua2}. Therefore, imposing the additional CP condition $\text{Im}[\tan z]=0$ ($\text{Im}[\tanh z]=0$) leads to the relations shown in table~\ref{condIm0R1st} (table~\ref{condIm0R2ndR3rd}) for $\OCI$ of the form $\OI$ ($\OII$ and $\OIII$). A straightforward conclusion can be drawn from these tables, namely, for the set of textures (R$_1$, Y$_{1,4}$, $\textbf{O}_{\text{I-III}}$), (R$_2$, Y$_{4}$, $\textbf{O}_{\text{II,III}}$) and (R$_3$, Y$_{1}$, $\textbf{O}_{\text{II,III}}$), we get $\alpha=k\pi$ or $(2k+1)\pi/2$, for $k\in\mathbb{Z}$. Also, when $\Mr$ is of the type R$_1$, the predictions are independent of the mass ratio $r_N$. As shown in table~\ref{condIm0R1st}, this may not be the case for the patterns R$_{2,3}$. Notice that, when there is a dependence on $r_N$, the results are invariant under the transformation $r_N\rightarrow 1/r_N$. Furthermore, the predictions for a combination (R$_{2}$,Y$_{i}$) are the same as the ones obtained for (R$_{3}$,Y$_{j}$), where Y$_{j}$ corresponds to Y$_{i}$ with interchanged columns. As it is apparent from table~\ref{condIm0R2ndR3rd}, the results are independent of the $r_N$ and $\xi$ values for $\OII$ and $\OIII$. We recall that, in these two cases, the condition $|\tanh z|<1$ must also be ensured. This implies further constraints among the relevant parameters,\footnote{The analytical expressions for these constraints are much more complicated than those presented in tables~\ref{condIm0R1st} and~\ref{condIm0R2ndR3rd} and, thus, we will not show them here. The numerical results obtained for the cases with $\OII$ and $\OIII$ fully take into account such constraints and will be presented in section~\ref{sec4}.} which may or may not be compatible with those of table~\ref{condIm0R2ndR3rd}. For illustration, let us consider the simplest combination (R$_1$,Y$_{1,4}$), for which the condition $|\tanh z|<1$ is equivalent to
\begin{align}
(\text{Y}_1,\OII),\,(\text{Y}_4,\OIII): \quad
&\left|\frac{\sqrt[4]{r_\nu}\,s_{12}}{t_{13}}\sin\alpha\right|<1\;,\quad\quad\;\,\text{NO}\,,\label{rel1}\\
&\left|\frac{1}{\sqrt[4]{1+r_\nu}\,t_{12}}\,\sin\alpha\right|<1\;,\;\;\,\text{IO}\,,\label{rel2}\\
(\text{Y}_1,\OIII),\,(\text{Y}_4,\OII): \quad
&\left|\frac{t_{13}}{\sqrt[4]{r_\nu}\,s_{12}}\sin\alpha\right|<1\;,\quad\quad\;\,\text{NO}\,,\label{rel3}\\
&\left|\sqrt[4]{1+r_\nu}\,t_{12}\,\sin\alpha\right|<1\;,\quad\text{IO}\,,\label{rel4}
\end{align}
depending on the form of $\OCI=\textbf{O}_{\text{II,III}}$. From table~\ref{datatable}, it is clear that the first two relations are incompatible with the experimental data, while the last two are always verified. Therefore, only the cases ($\text{Y}_1,\OIII$) and ($\text{Y}_4,\OII$) are potentially viable. A complete numerical analysis of the constraints for each of the considered patterns of $\Ynu$, $\Mr$ and $\OCI$ will be presented in section~\ref{sec4}.

Given that, for degenerate RH neutrinos and $(a,b)=\pm(1,1)$, the matrix $\OCI$ is constrained to be real and of the type $\OI$ (see discussion after eq.~\eqref{ONconst2}), the results in table~\ref{condIm0R1st} are valid for all combinations $(\text{R}_1,\text{Y}_{1-6})$. If, instead, $\Mr$ is of the type R$_4$, the texture-zero conditions~\eqref{Ynua1} and~\eqref{Ynua2} imply $\tan z=\pm i$, which is incompatible with the requirement of $\OCI$ being real. Therefore, all combinations (R$_{4}$,Y$_{1-6}$) are excluded for $(a,b)=\pm(1,1)$.

To conclude this section, let us briefly comment on the compatibility of two texture-zero patterns for $\Ynu$ in the present framework (see eq.~\eqref{ynupatterns2}). For the NO case, none of such textures is compatible with neutrino data, even in the absence of the CP symmetry~\cite{Barreiros:2018ndn}. This remains true for IO when $\Mr$ is of the type R$_1$, $r_N \neq 1$ and the CP symmetry is imposed, as can be seen by combining the results in table~\ref{condIm0R1st} or~\ref{condIm0R2ndR3rd} for two different patterns of $\Ynu$ with zeros in distinct columns~\cite{Barreiros:2018ndn}. For instance, when $\Mr$ is diagonal and $\Ynu$ is of the form T$_1$, the low-energy constraints are those coming from Y$_1$ and Y$_5$ simultaneously. As indicated in the second and fourth rows of table~\ref{condIm0R1st}, this corresponds to
\begin{align}
\sin\alpha=0\,,
\end{align}
and
\begin{align}
\tan\alpha=\frac{4\sin(2\theta_{23}) s_{13}\sin\delta}{\sin(2\theta_{12})[1+3\cos(2\theta_{23})+2\cos(2\theta_{13})s_{23}^2]+4\cos(2\theta_{12})\sin(2\theta_{23})s_{13}\cos\delta}\,,			
\end{align}
implying $(\alpha,\delta)=(k\pi,k\pi)$, with $k\in \mathbb{Z}$. These solutions for T$_1$ are not compatible with low-energy neutrino data~\cite{Barreiros:2018ndn}. Proceeding in the same way for the remaining patterns with $M_1\neq M_2$, and $M_1=M_2$ with $(a,b)=\pm(1,1)$, it can be shown that two texture zeros in $\Ynu$ are not compatible with neutrino oscillation data in the presence of a remnant CP symmetry, if $\Mr$ is of the type R$_1$.

When $r_N\neq1$ and $\Mr$ is nondiagonal (R$_2$ or R$_3$), the compatibility of the two texture zeros in $\Ynu$ with the CP symmetry calls for a more thorough analysis. The possible cases are (R$_2$,T$_{3,4,5,6}$) and (R$_3$,T$_{1,2,3,6}$) which are viable at 3$\sigma$ for IO when the CP symmetry is not imposed~\cite{Barreiros:2018ndn}. For instance, let us consider the set (R$_2$,T$_3$). In this case, the low-energy constraints are those coming from imposing the one texture zero conditions of Y$_2$ and Y$_6$ simultaneously, together with the CP symmetry. The  texture zero in Y$_2$ automatically implies the low-energy relation $(\Mnu)_{22}=0$, predicting the CP-violating phases $\alpha$ and $\delta$ (see table~\ref{tabR4}). On the other hand, the zero in Y$_6$ determines the $z$ parameter, which must be real due to the CP symmetry. This leads to the low energy relation in the last column of table~\ref{condIm0R1st} for $\OI$ or table~\ref{condIm0R2ndR3rd} for $\OCI_\text{II,III}$. In the case of $\OI$, the three low-energy constraints determine the ratio $r_N$. The cases with $\OII$ and $\OIII$ are somehow more restrictive due to their independence on $r_N$. A complete numerical analysis for each of the possible two texture zero patterns of $\Ynu$ and $\OCI$ will be presented in the end of section~\ref{sec4}, for R$_2$ and R$_3$.

\section{Parameter space analysis}
\label{sec4}
In this section, we will investigate the compatibility with data of the 2RHNSM with maximally-restricted texture zeros in the presence of a remnant CP symmetry. Our analysis is focused on the predictions for leptonic CP violation and $0\nu\beta\beta$ decay. Having excluded some of the two and one-texture zero cases in the previous section, we will start by analysing the viability of the one-texture zero sets $(\text{R}_{1-3},\text{Y}_{1-6})$, for nondegenerate RH neutrinos, both in the NO and IO cases.\footnote{The one-texture zero results obtained in this section for $\Mr$ of the type R$_1$ (with $M_1 \neq M_2$) and $\OCI=\OI$ are also valid for $\Mr$ of the type R$_1$ with $M_1=M_2$ and $(a,b)=\pm(1,1)$.} Later on, we will check the agreement with data of the two-texture zero cases (R$_2$,T$_{3,4,5,6}$) and (R$_3$,T$_{1,2,3,6}$) with IO, also for nondegenerate RH neutrinos.
\begin{figure}[t]
	\centering
	\includegraphics[trim={0.8cm 0.2cm 0.8cm 0.1cm},clip,scale=1.02]{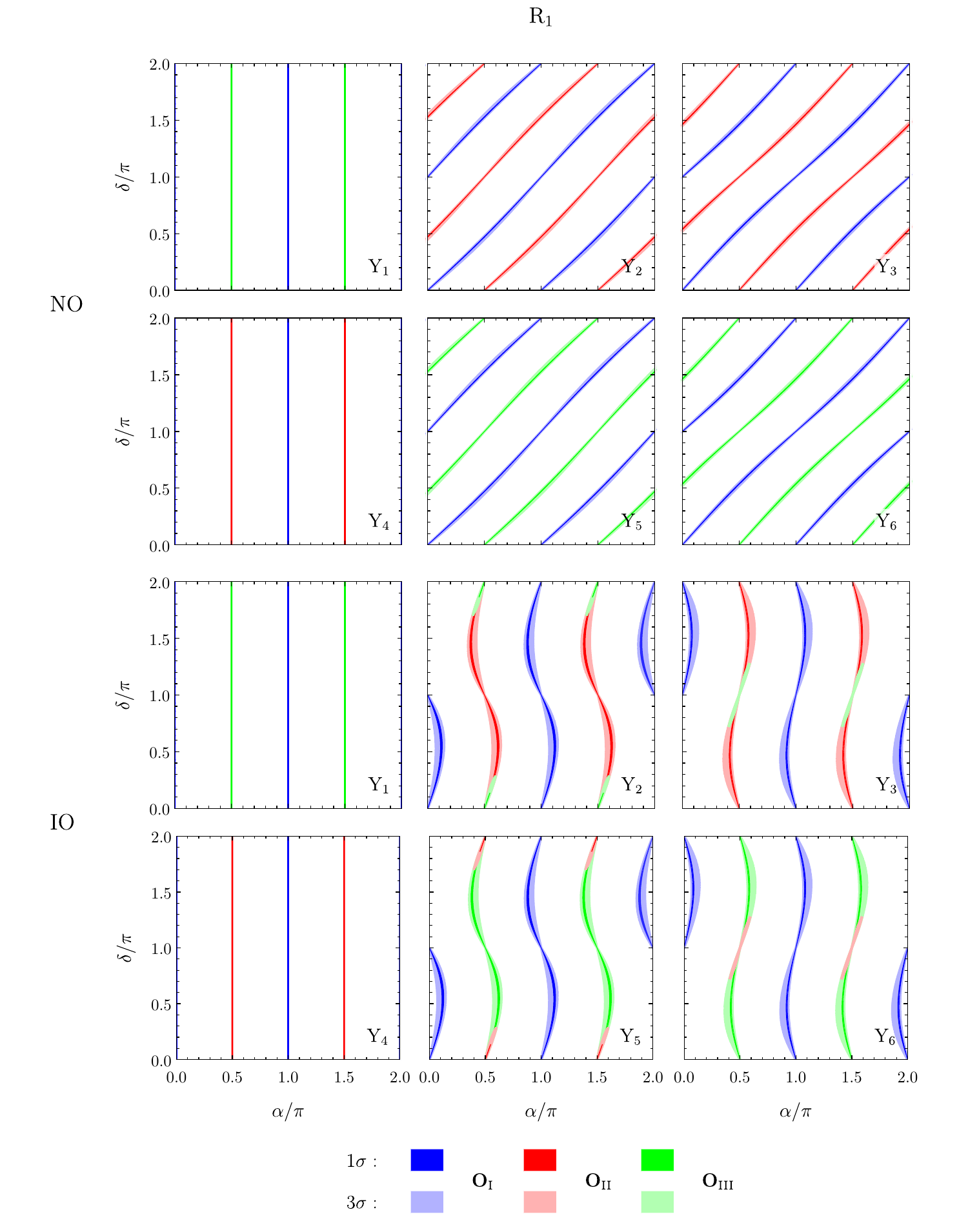}
	\caption{Allowed regions in the ($\alpha$, $\delta$) plane for all (R$_1$,Y$_{1-6}$) combinations, both for NO (upper plots) and IO (lower plots). The light (dark) coloured regions were obtained varying $\theta_{ij}$ and $\Delta m_{ij}^2$ within the $3\sigma$ ($1\sigma$) experimental ranges given in table~\ref{datatable}. In blue, red and green we show the results for the cases in which $\OCI$ is of the type $\OI$, $\OII$ and $\OIII$, respectively.}
	\label{neutrinoMR1}
\end{figure}
\begin{figure}[t]
	\centering
	\includegraphics[trim={1.4cm 0.2cm 1.4cm 0.1cm},clip,scale=0.99]{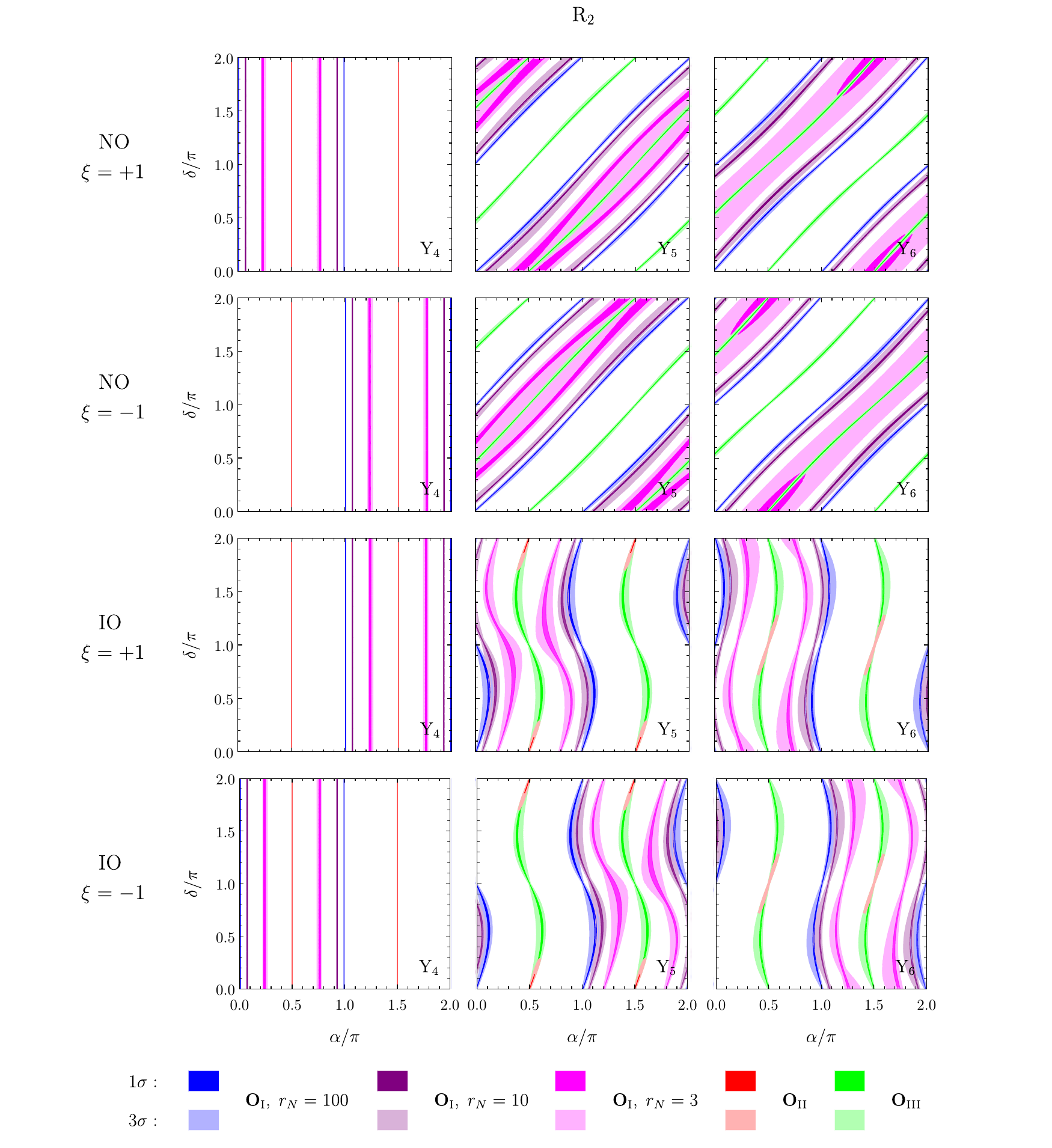}
	\caption{Allowed regions in the ($\alpha$, $\delta$) plane for all (R$_2$,Y$_{4-6}$) combinations, both for NO (upper plots) and IO (lower plots), and $\xi=\pm1$. The light (dark) coloured regions were obtained varying $\theta_{ij}$ and $\Delta m_{ij}^2$ within the $3\sigma$ ($1\sigma$) experimental ranges given in table~\ref{datatable}. In blue, purple, magenta, red and green we show the results for the cases in which $\OCI$ is of the type $\OI$ for $r_N=100$, $\OI$ for $r_N=10$, $\OI$ for $r_N=3$, $\OII$ and $\OIII$, respectively.}
	\label{neutrinoMR2}
\end{figure}

\subsection{Leptonic CP violation }
Let us start by analysing the compatibility with data of the one-zero $\Ynu$ patterns in eq.~\eqref{ynupatterns} in the presence of a CP symmetry. Taking into account the data of table~\ref{condIm0R1st}, and of table~\ref{condIm0R2ndR3rd} combined with the condition $|\tanh z|<1$, we now present the allowed regions in the $(\alpha,\delta)$ plane for each $\Ynu$ and $\Mr$ textures, considering the $3\sigma$ and $1\sigma$ experimental ranges for the neutrino mixing angles and mass-squared differences (see table~\ref{datatable}). In figure~\ref{neutrinoMR1}, we show the results for all (R$_1$,Y$_{1-6}$) combinations. The two upper (lower) panels correspond to the case of NO (IO) neutrino mass spectrum. For a given texture Y and depending on the type of $\OCI$ imposed by the CP symmetry (see table~\ref{Rforms}), we obtain different compatibility regions, presented in blue, red and green for $\OI$, $\OII$ and $\OIII$, respectively. As already mentioned, $\alpha=k\pi$ and $(2k+1)\pi/2$ for textures Y$_{1,4}$ with $\OI$ and $\textbf{O}_{\text{II,III}}$, respectively, being this result independent of the remaining low-energy parameters (see the first two rows of tables~\ref{condIm0R1st} and \ref{condIm0R2ndR3rd}). In the case of NO with Y$_{2,3,5,6}$, the approximate relations $\delta \simeq \alpha + k \pi$ and $\delta \simeq \alpha + (2k-1) \pi/2$ hold for $\textbf{O}_{\text{I}}$ and $\textbf{O}_{\text{II,III}}$, respectively. Instead, in the IO case, $\alpha$ oscillates around $\alpha=k\pi$ or $(2k-1)\pi/2$ if $\OCI$ takes the form $\OI$ or $\textbf{O}_{\text{II,III}}$, respectively, for unconstrained $\delta$.

\begin{figure}[t]
	\centering
	\includegraphics[trim={1.4cm 0.3cm 1.4cm 0.1cm},clip,scale=1.0]{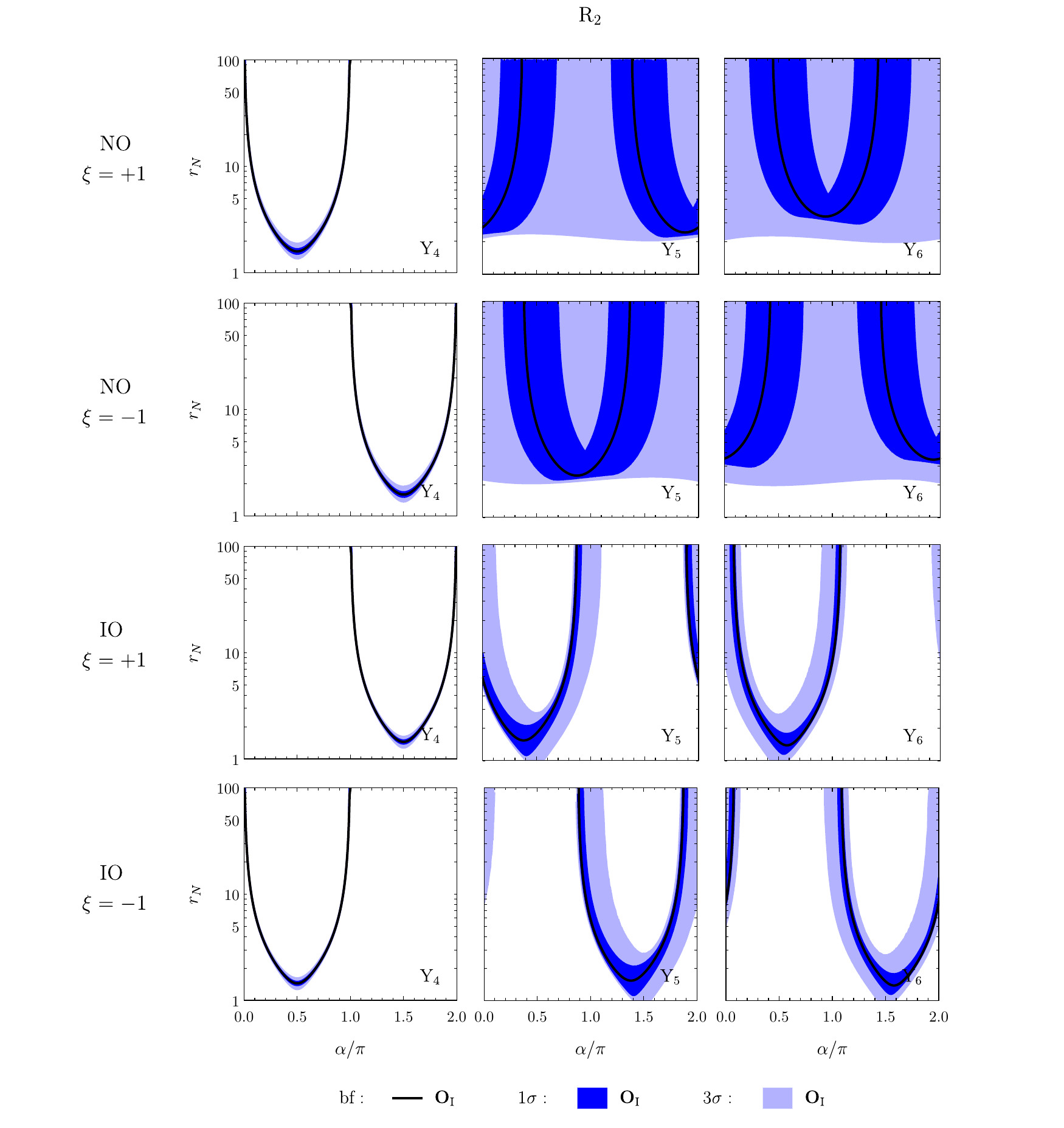}
	\caption{Allowed regions in the ($\alpha$, $r_N$) plane for all (R$_2$,Y$_{4-6}$) combinations, both for NO (upper plots) and IO (lower plots), $\xi=\pm1$, and $\OCI$ of the type $\OI$. The light (dark) coloured regions were obtained varying $\theta_{ij}$, $\Delta m_{ij}^2$ and $\delta$ within the $3\sigma$ ($1\sigma$) experimental ranges given in table~\ref{datatable}. The black solid lines are the ($\alpha$, $r_N$) curves computed with the best-fit values for neutrino data. Here, we only present the case where $\OCI$ is of the type $\OI$, since it is the only one that depends on $r_N$.}
	\label{neutrinoMR2alpharN}
\end{figure}

For a non-diagonal $\Mr$ of the form R$_2$, only the cases Y$_{4,5,6}$ lead to constraints coming from imposing $z$ real, which are indicated in tables~\ref{condIm0R1st} and~\ref{condIm0R2ndR3rd}. For these textures, the allowed regions are depicted in figure~\ref{neutrinoMR2}, for both NO (upper plots) and IO (lower plots),\footnote{We will not show the results for R$_3$ since the predictions for the patterns Y$_1$, Y$_2$ and Y$_3$ are the same as those for Y$_4$, Y$_5$ and Y$_6$, respectively, taking $\Mr$ of the form R$_2$ (see figures~\ref{neutrinoMR2} and \ref{neutrinoMR2alpharN}). \label{foot4}} considering $\xi=\pm 1$. Since for $\OI$ the results depend on the mass ratio $r_N$, we consider the representative values $r_N=3,10,100$ (blue, purple and pink regions, respectively). Notice that, for $r_N\gg 1$, the regions are very similar to those obtained for diagonal $\Mr$, where there is no dependence on $r_N$ (cf. table~\ref{condIm0R1st}). For Y$_4$, this is easily understood taking the limit $r_N \gg 1$ in the expressions of table~\ref{condIm0R1st}, while for Y$_{5,6}$ the complete expressions would be needed. Moreover, from comparison of the results in figures~\ref{neutrinoMR1} and \ref{neutrinoMR2} for $\OII$ and $\OIII$ (in red and green), it can be seen that the same allowed regions appear for R$_{1}$ and R$_{2}$, which is explained by the results given in table~\ref{condIm0R2ndR3rd}. Up to now, we have not taken into account the experimental constraints on the CP-violating phase $\delta$, due to its poor statistical significance at present. Varying this phase within the experimental range given in table~\ref{datatable}, we can study the allowed regions in the space of the only two free parameters in the model, namely, $\alpha$ and $r_N$. The results presented in figure~\ref{neutrinoMR2alpharN} show that, in most cases, $r_N$ and $\alpha$ are strongly correlated and nontrivial bounds on $r_N$ can be set. Moreover, for the most part, the allowed range for $r_N$ may be very narrow, depending the value of $\alpha$. The remaining R$_2$ (R$_3$) cases with Y$_{2,3}$ (Y$_{5,6}$), which correspond to textures D and F in table~\ref{tabR4}, are also viable at 3$\sigma$ for IO, independently of $r_N$, $\xi$ and the form of $\OCI$~\cite{Barreiros:2018ndn}.
\begin{figure}[t]
	\centering
	\includegraphics[scale=0.95]{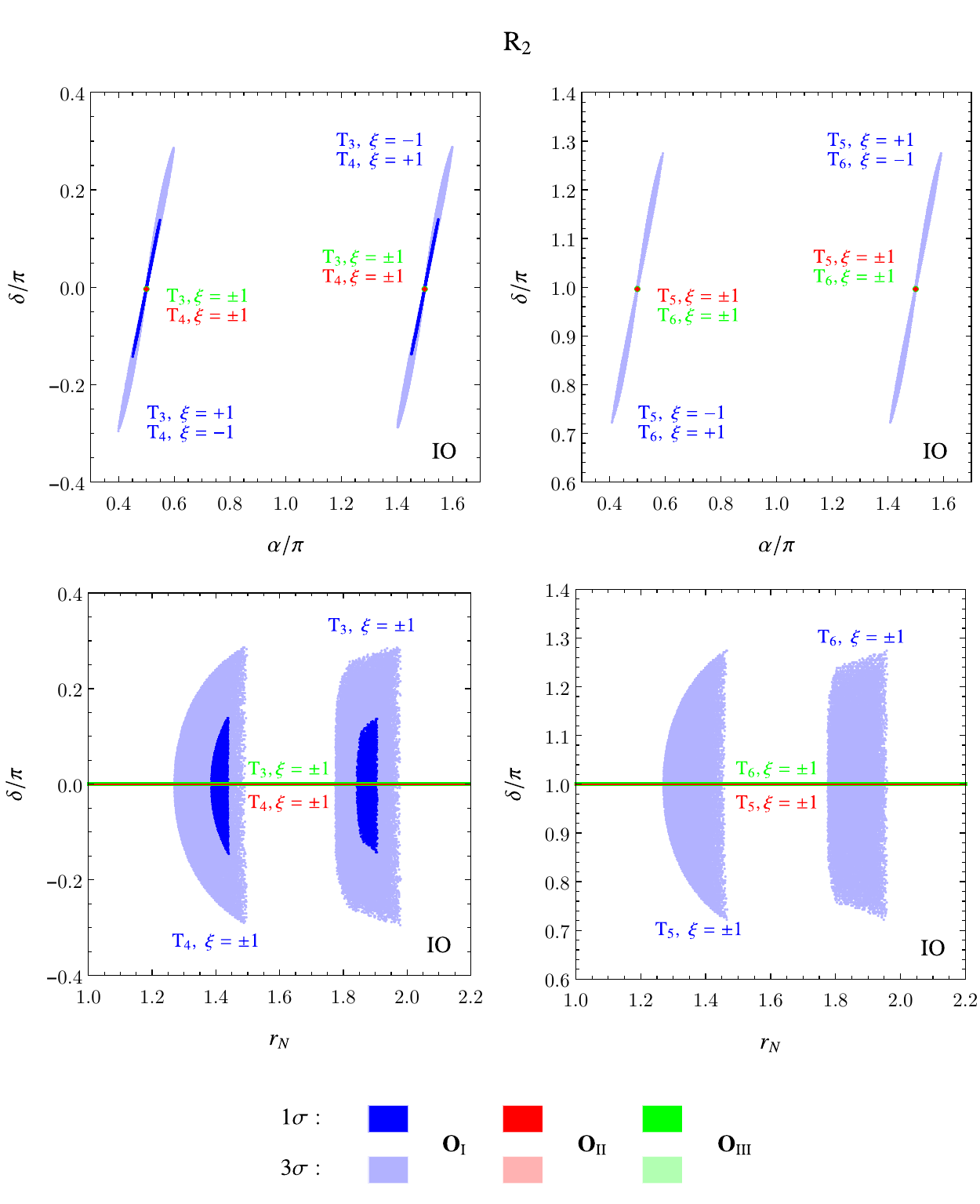}
	\caption{Allowed regions in the ($\alpha$, $\delta$) and ($r_N$,$\delta$) planes (upper and lower plots, respectively) for the (R$_2$, T$_{3,4}$, $\xi=\pm1$) and (R$_2$, T$_{5,6}$, $\xi=\pm1$) combinations (left and right plots, respectively), in the case of IO and $\OCI$ of the types $\mathbf{O}_\text{I-III}$. The light (dark) colored regions were obtained varying $\theta_{ij}$ and $\Delta m_{ij}^2$ within the $3\sigma$ ($1\sigma$) experimental ranges given in table~\ref{datatable}.}
	\label{R2deltaalpharn2TZ}
\end{figure}

Having concluded that all $(\text{R}_{1,2,3},\text{Y}_{1-6})$ are viable for at least one form of the matrix $\OCI$, let us now analyse the combinations (R$_2$,T$_{3,4,5,6}$) and (R$_3$,T$_{1,2,3,6}$), with two-texture zeros in $\Ynu$, for IO. Taking into account the constraints for $\alpha$ and $\delta$ given in tables~\ref{tabR4} and~\ref{condIm0R1st} (or table~\ref{condIm0R2ndR3rd} combined with the condition $|\tanh z|<1$), we present in figure~\ref{R2deltaalpharn2TZ} the allowed regions in the $(\alpha,\delta)$ and $(r_N,\delta)$ planes, for textures (R$_2$,T$_{3,4,5,6}$).\footnote{Since the predictions for cases with R$_3$ are the same as those with R$_2$, by exchange of the two columns in the T patterns, we will only present the results for (R$_2$,T$_{3,4,5,6}$) in this section.} As discussed in the end of section~\ref{sec3}, for $\OI$ (in blue) the same $(\alpha,\delta)$ regions are obtained as for textures D and F, depending on the sign of $\xi$. However, the presence of the CP symmetry (see table~\ref{condIm0R1st}) implies $r_N \simeq 1.4$ for textures T$_4$ and T$_5$, and $r_N \simeq 1.9$ for T$_3$ and T$_6$. Notice that for the cases with $\OII$ (in red) and $\OIII$ (in green), there is no dependence of the low-energy constraints on $r_N$ and  $\xi$, being the compatibility regions more restrictive. For textures T$_{3,4}$ (T$_{5,6}$), only the solutions $\delta=0$ ($\delta=\pi$) with $\alpha=\pi/2,3\pi/2$ are compatible with data.

\subsection{Neutrinoless double beta decay}
\begin{figure}[t]
	\centering
	\includegraphics[scale=0.95]{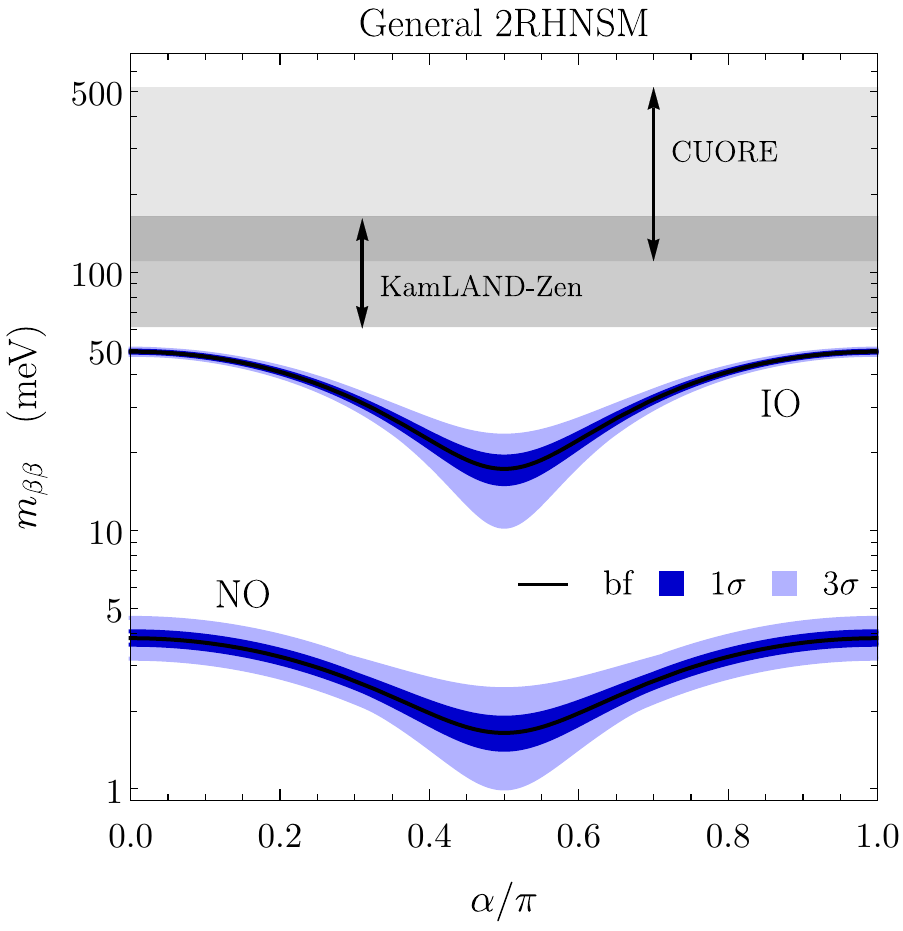}
	\caption{$m_{\beta\beta}$ predictions as a function of $\alpha$ for the general 2RHNSM (NO and IO). The dark (light) blue regions were obtained using eqs.~\eqref{NOmbb} and~\eqref{IOmbb}, and varying $\theta_{ij}$ and $\Delta m_{ij}^2$ within the $3\sigma$ ($1\sigma$) experimental ranges given in table~\ref{datatable}. The solid black curves correspond to $m_{\beta\beta}$ computed with the best-fit values for the neutrino parameters. For illustration, the present CUORE~\cite{Alduino:2017ehq} and KamLAND-Zen~\cite{Asakura:2015ajs} upper-limit intervals on $m_{\beta\beta}$ are also shown.}
	\label{mbb2RHNSM}
\end{figure}

When the only relevant contributions to $0\nu\beta\beta$ decays are due to the exchange of light neutrinos,\footnote{In this work, we consider that the heavy neutrino masses (Dirac Yukawa couplings) are large (small) enough to neglect the $0\nu\beta\beta$ contributions induced by the exchange of the heavy Majorana neutrinos~\cite{Ibarra:2011xn}.} the corresponding rates are sensitive to the effective mass
\begin{align}
	m_{\beta\beta}=\left|\sum_{i=1}^{3}m_i(\U^*_{1i})^2\right|=\left|m_1 c_{12}^2c_{13}^2+m_2 c_ {13}^2s_{12}^2e^{-2i\alpha_{21}}+m_3s_{13}^2e^{-2i\alpha_{31}}\right|\,.
\end{align}
In the general 2RHNSM, i.e. without imposing any constraint in the model parameters, one has
\begin{align}
\text{NO:}\;\;	m_{\beta\beta}=&\sqrt{\left|\dmatm\right|}\left|\sqrt{r_\nu}\,c_ {13}^2s_{12}^2\,e^{-2i\alpha}+ s_{13}^2\right|\,, \label{NOmbb}\\
\text{IO:}\;\;	m_{\beta\beta}=&\sqrt{\left|\dmatm\right|}\,c_{13}^2\left|c_{12}^2+\sqrt{1+r_\nu}\,s_{12}^2\,e^{-2i\alpha}\right|\label{IOmbb}\,,
\end{align}
for NO and IO. These expressions, together with the experimental ranges for the neutrino parameters given in table~\ref{datatable}, lead to the allowed regions in the $(\alpha,m_{\beta\beta})$ plane presented in figure~\ref{mbb2RHNSM}. To illustrate the present experimental sensitivity of $0\nu\beta\beta$ experiments, the upper-limit intervals for $m_{\beta\beta}$ coming from the CUORE~\cite{Alduino:2017ehq} and KamLAND-Zen~\cite{Asakura:2015ajs} experiments are also shown. 
Taking the $1\sigma$ and $3\sigma$ ranges for current neutrino oscillation data, the following $m_{\beta\beta}$ ranges are obtained:
\begin{align}
	\text{NO:}\;\;	&\,[1.4 \,(1.0),4.1 \,(4.6)]\;\text{meV}, \label{NOmbbrange}\\
	\text{IO:}\;\;	&\, [19.3 \,(10.3),48.6\, (50.4)]\;\text{meV}\label{IOmbbrange},
\end{align}
for the general 2RHNSM, where the parenthesis indicate the $3\sigma$ intervals. The sensitivity of near-future experiments like KamLAND-Zen, CUORE, SNO+, NEXT and LEGEND, partially covers the IO ranges. On the other hand, the NO scenario will only be probed by new-generation projects (discussions on present results and prospects for neutrinoless double beta decay experiments can be found in refs.~\cite{DellOro:2016tmg,Vergados:2016hso,Giuliani:neutrino}). 
\begin{figure}[]
	\begin{tabular}{c}
		\includegraphics[scale=0.095]{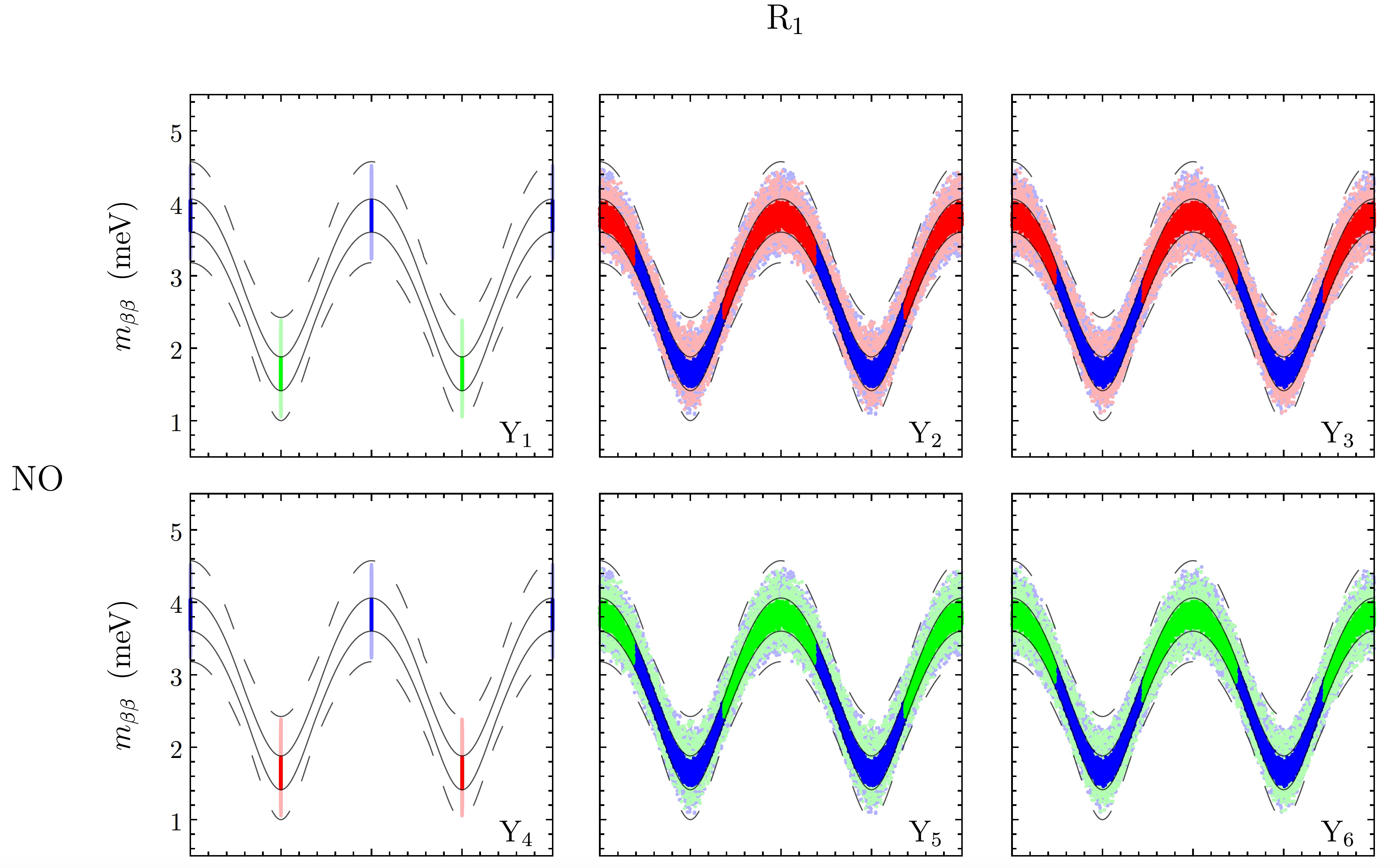}\\[0.1cm]
		\includegraphics[scale=0.095]{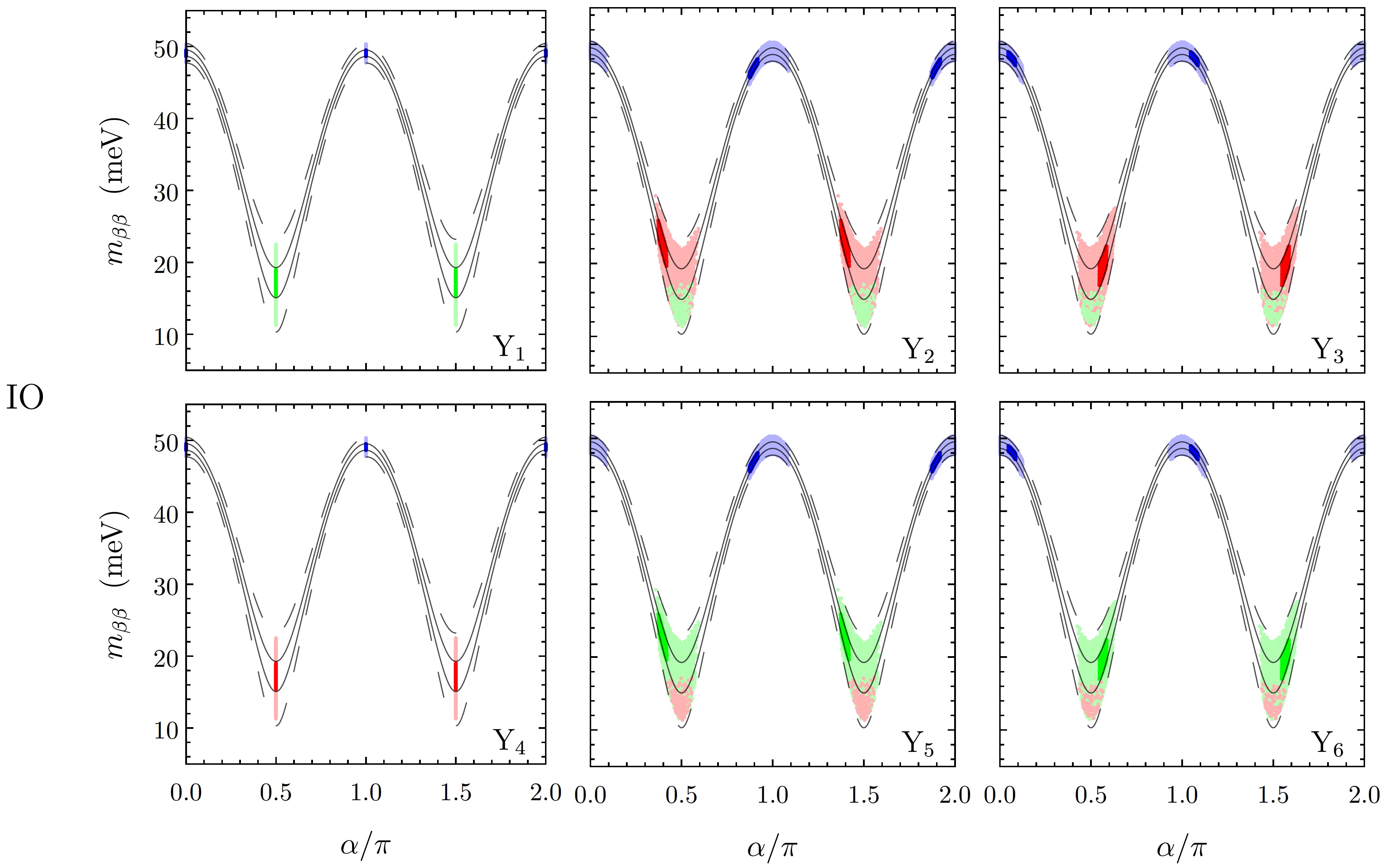}\\[0.2cm]
		\includegraphics[scale=0.095]{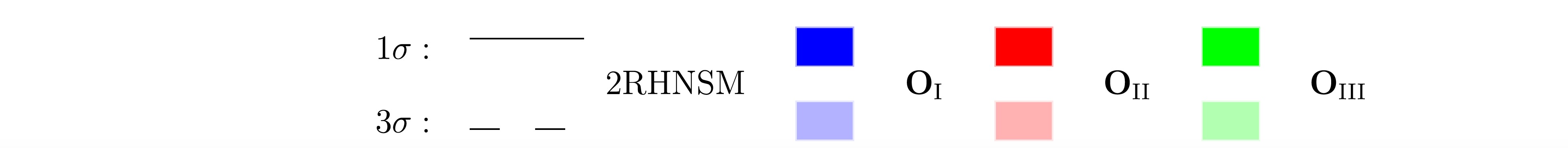}
	\end{tabular}
	\caption{Allowed regions in the ($\alpha$, $m_{\beta\beta}$) plane for all (R$_1$,Y$_{1-6}$) combinations, both for NO (upper plots) and IO (lower plots). The light (dark) coloured points were obtained varying $\theta_{ij}$, $\Delta m_{ij}^2$ and $\delta$ within the $3\sigma$ ($1\sigma$) experimental ranges given in table~\ref{datatable}. In blue, red and green we show the results for the cases in which $\OCI$ is of the type $\OI$, $\OII$ and $\OIII$, respectively. The solid (dashed) lines are the contours of the $1\sigma$ ($3\sigma$) $m_{\beta\beta}$ regions allowed by the general 2RHNSM previously shown in figure~\ref{mbb2RHNSM}.}
	\label{alphambbR1}
\end{figure}
\begin{figure}[]
	\begin{tabular}{c}
		\includegraphics[scale=0.105]{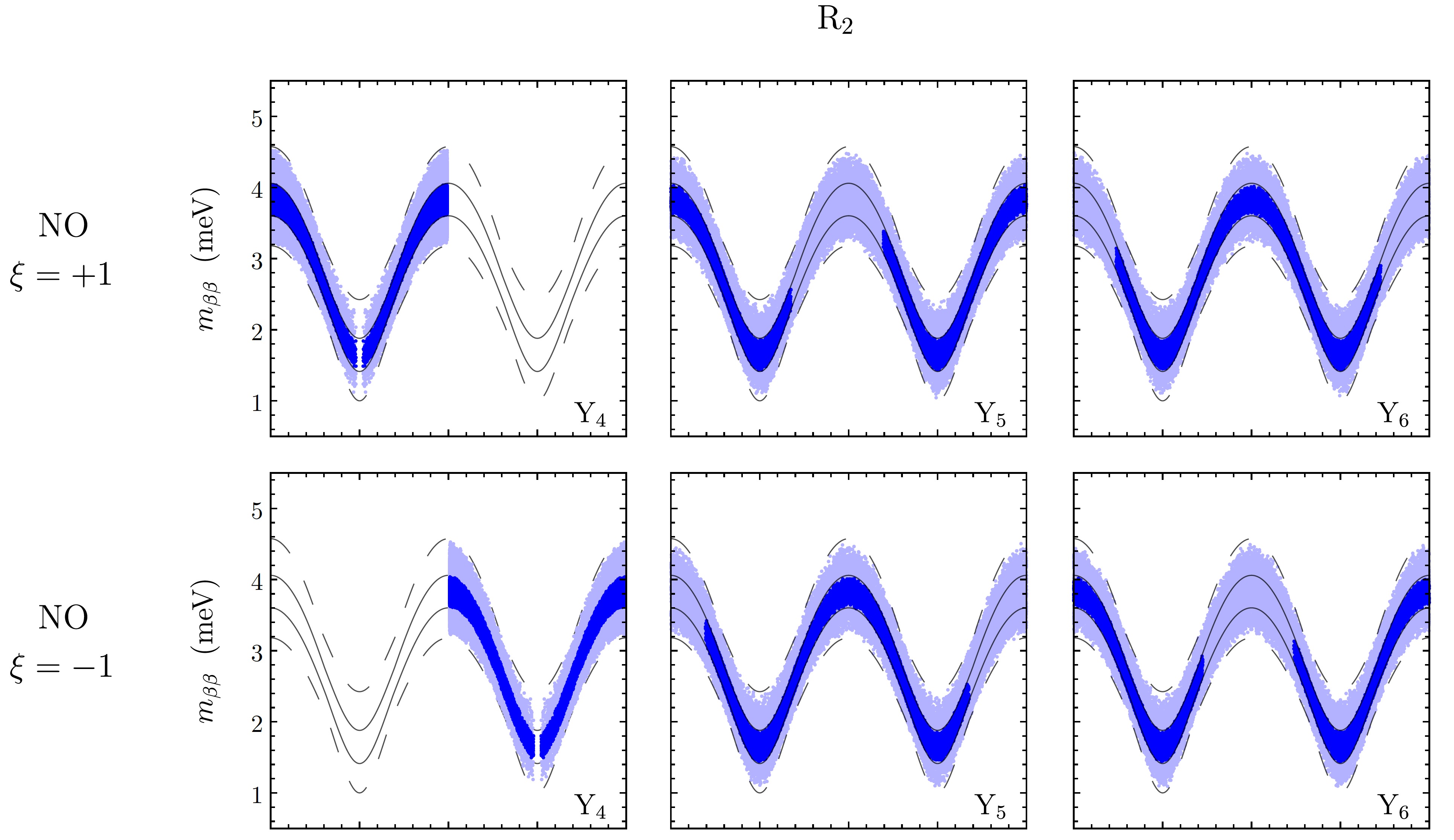}\\[0.1cm]
		\includegraphics[scale=0.105]{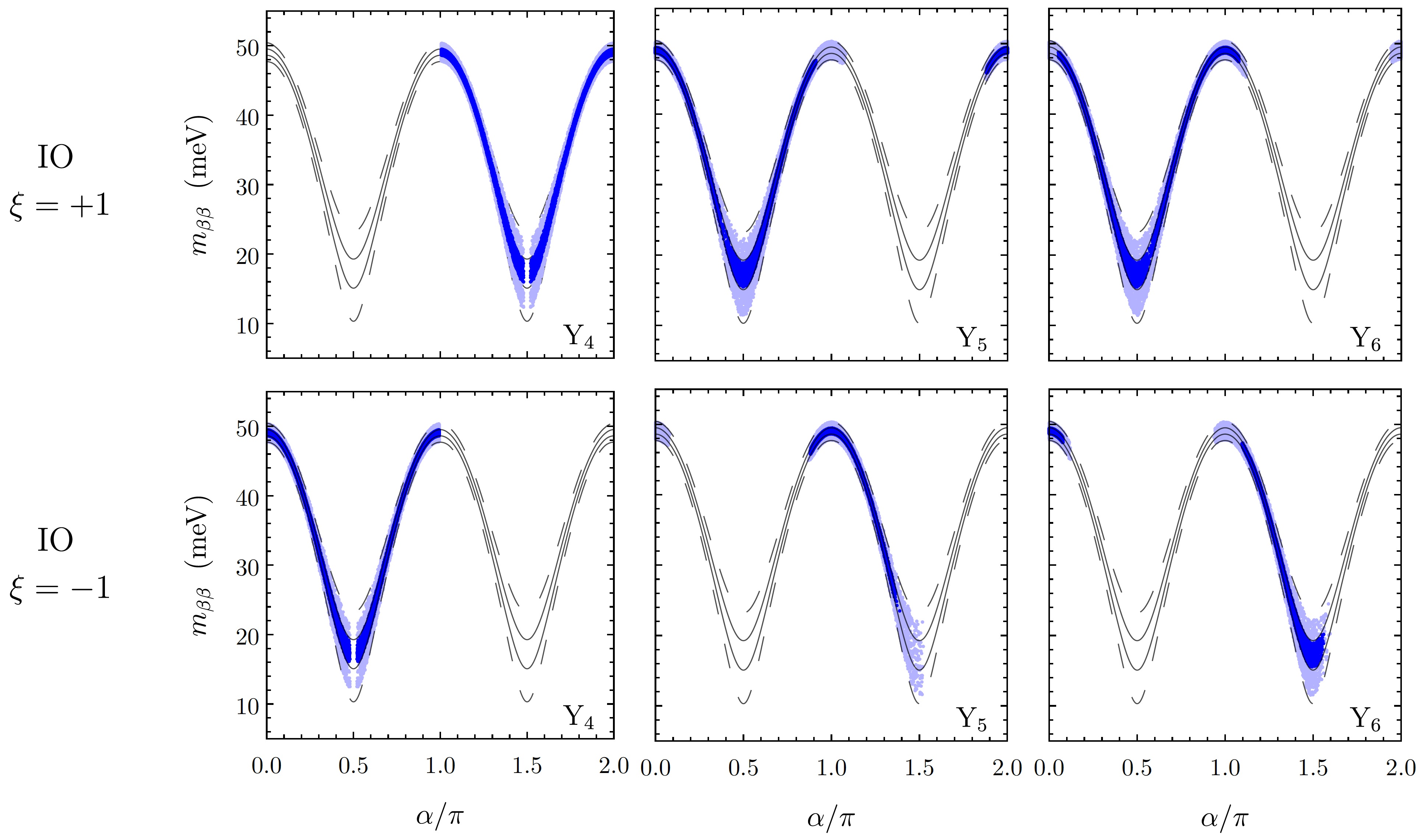}\\[0.2cm]
		\includegraphics[scale=0.105]{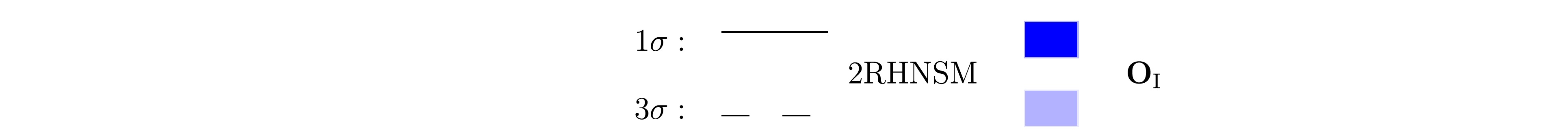}
	\end{tabular}
	\caption{Allowed regions in the ($\alpha$, $m_{\beta\beta}$) plane for all (R$_2$,Y$_{4-6}$) combinations, both for NO (upper plots) and IO (lower plots), $\xi=\pm1$, and $\OCI$ of the type $\OI$. The light (dark) blue points were obtained varying $\theta_{ij}$, $\Delta m_{ij}^2$ and $\delta$ within the $3\sigma$ ($1\sigma$) experimental ranges given in table~\ref{datatable}. The solid (dashed) lines are the contours of the $1\sigma$ ($3\sigma$) $m_{\beta\beta}$ regions allowed by the general 2RHNSM previously shown in figure~\ref{mbb2RHNSM}. The compatibility regions for $\OII$ and $\OIII$ are the same as those obtained in figure~\ref{alphambbR1}.}
	\label{alphambbR2}
\end{figure}

\begin{figure}[t]
	\centering
	\includegraphics[scale=1.0]{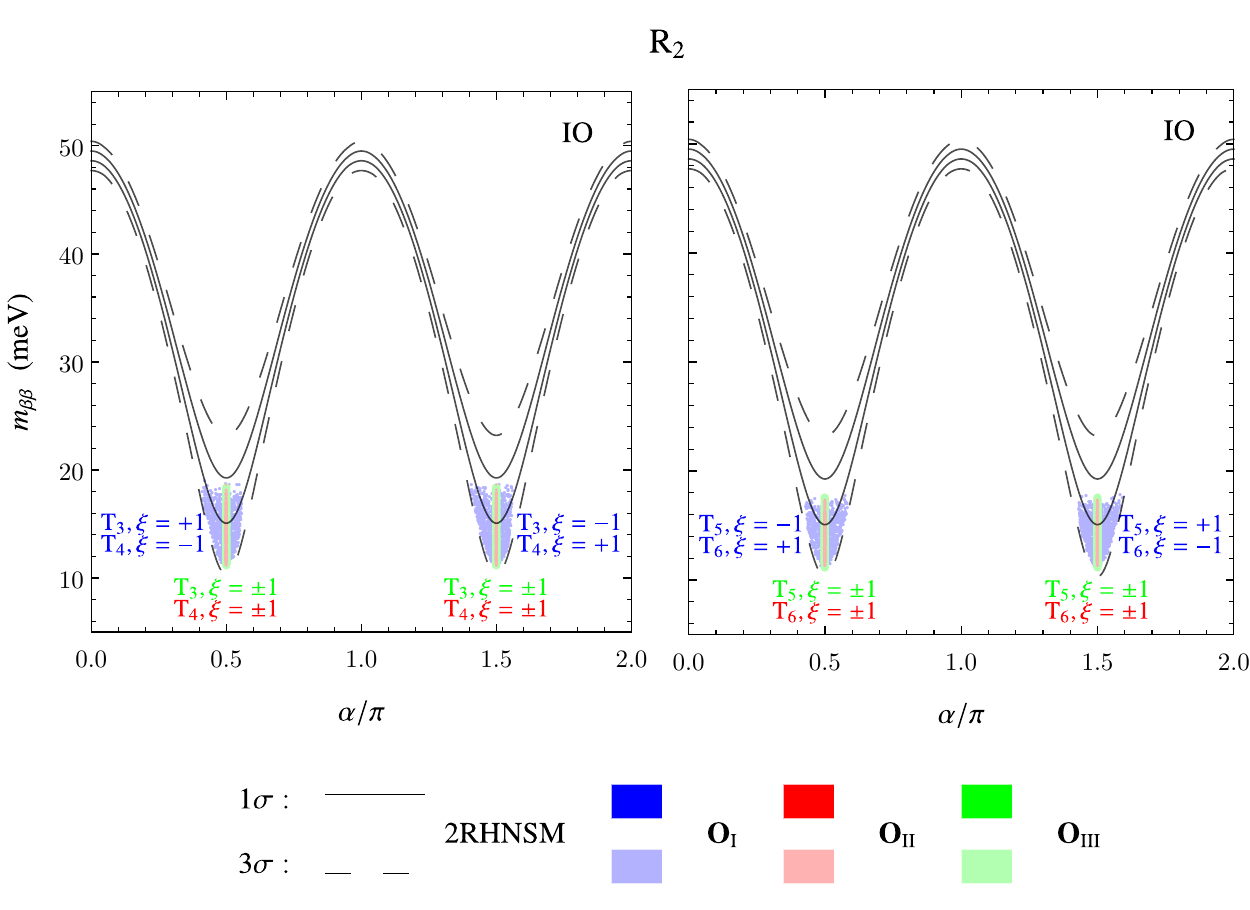}
	\caption{Allowed regions in the ($\alpha$,$m_{\beta\beta}$) plane for the (R$_2$, T$_{3,4}$, $\xi=\pm1$) and (R$_2$, T$_{5,6}$, $\xi=\pm1$) combinations (left and right plots, respectively), in the case of IO and $\OCI$ of the types $\mathbf{O}_\text{I-III}$. The light (dark) colored regions were obtained varying $\theta_{ij}$, $\Delta m_{ij}^2$ and $\delta$ within the $3\sigma$ ($1\sigma$) experimental ranges given in table~\ref{datatable}. The solid (dashed) lines are the contours of the $1\sigma$ ($3\sigma$) $m_{\beta\beta}$ regions allowed by the general 2RHNSM previously shown in figure~\ref{mbb2RHNSM}.}
	\label{R2alphambb2TZ}
\end{figure}

Given the above results, we will now analyse the impact on $m_{\beta\beta}$ of the previously discussed texture-zero patterns and remnant CP symmetry. In figures~\ref{alphambbR1} and~\ref{alphambbR2}, we show the predictions for $m_{\beta\beta}$ when the constraints given in tables~\ref{condIm0R1st} and \ref{condIm0R2ndR3rd} (and the condition $|\tanh z|<1 $) are imposed on $\alpha$ and current neutrino data of table~\ref{datatable} is used. The blue, red and green points correspond to the cases in which $\OCI$ is of the type $\OI$, $\OII$ and $\OIII$, respectively. Furthermore, darker (lighter) colours stand for the predictions obtained taking the $1\sigma$ ($3\sigma$) intervals for all parameters given in table~\ref{datatable}. The results for R$_1$ are presented in figure~\ref{alphambbR1} for the $\Ynu$ textures Y$_{1-6}$, while in figure~\ref{alphambbR2} the results for R$_2$ are given for Y$_{4-6}$. In the latter figure, $\xi=\pm 1$ is considered, and only the case $\OCI=\OI$ is shown, since $\OII$ and $\OIII$ lead to the same predictions as those with R$_1$ (see figure~\ref{alphambbR1}). Moreover, $r_N$ varies within the compatibility regions shown in figure~\ref{neutrinoMR2alpharN}. For comparison, we include in both figures the delimiting lines of the NO and IO regions of figure~\ref{mbb2RHNSM}, valid for the general 2RHNSM. As commented in footnote~\ref{foot4}, the results for R$_3$ are related to those of R$_2$. The $m_{\beta\beta}$ regions for R$_2$ become even more restricted if textures T$_{3-6}$ are considered, as can be concluded from figure~\ref{R2alphambb2TZ}, where only the viable IO cases are shown. Here the results for $\OCI_\text{II,III}$ are independent of $r_N$ and $\xi$, while the results for $\OI$ correspond to a very narrow region of $r_N$ for $\alpha\simeq\pi/2$ and $\alpha\simeq3\pi/2$, which lead to $m_{\beta\beta}$ values in the small range $\sim[10,18]$ meV.

When compared to the general 2RHNSM, the results of figures~\ref{alphambbR1}, \ref{alphambbR2} and~\ref{R2alphambb2TZ} show that, imposing texture zeros in $\Ynu$ and $\Mr$, together with a remnant CP symmetry, can have a profound impact on the $m_{\beta\beta}$ predictions. In fact, a mild improvement of the current experimental limits on this parameter would disfavour some of the pattern combinations of $\Mr$ and $\Ynu$ considered in this work. This would be the case for the IO combinations (R$_1$,Y$_{1-6}$), with $\OCI=\OI$.

We conclude this section by remarking that our results are valid in the basis in which the charged-lepton Yukawa matrix $\Yl$ is diagonal. When $\Yl$ is non-diagonal (but still contains six texture zeros), the analysis can be performed following the same procedure of ref.~\cite{Barreiros:2018ndn}, i.e., by applying column and/or row permutations on the textures considered here for the diagonal case.
\begin{table}[t!]
	\centering
	\setlength\extrarowheight{2pt}
	\begin{tabular}{ccc}
		\hline\hline
		\multicolumn{3}{c}{Degenerate RH neutrinos ($r_N = 1$)
		}\\ 
		\hline\hline\\[-0.55cm]
		\multicolumn{3}{c}{$\Xhnu$ with $(a,b)=\pm(1,-1)$}\\ 
		\hline 
		$\Mr$&$\Ynu$&Compatibility (at least $3\sigma$)\\
		\hline
		\multirow{3}{*}{R$_1$}&T$_{1,2,4,5}$&IO\\
		\cline{2-3}
		&T$_{3,6}$&\xmark\\	
			\cline{2-3}
		&Y$_{1-6}$&NO, IO\\
		\hline
		\multirow{3}{*}{R$_4$}&T$_{1-6}$&\xmark\\
		\cline{2-3}
		&Y$_{1,4}$&\xmark\\		
		\cline{2-3}
		&Y$_{2,5,3,6}$&IO\\	
		\hline\hline\\[-0.55cm]
		\multicolumn{3}{c}{$\Xhnu$ with $(a,b)=\pm(1,1)$}\\ 
		\hline 	
		$\Mr$&$\Ynu$&Compatibility (at least $3\sigma$)\\
		\hline
		\multirow{2}{*}{R$_1$}&T$_{1-6}$&\xmark\\
		\cline{2-3}
		&Y$_{1-6}$&NO, IO\\	
		\hline
		\multirow{2}{*}{R$_4$}&T$_{1-6}$&\xmark\\
		\cline{2-3}
		&Y$_{1-6}$&\xmark\\					
		\hline\hline
	\end{tabular}
	\caption{Compatibility summary for the 2RHNSM in presence of the remnant CP symmetry defined in eqs.~\eqref{CPnuL} and~\eqref{CPnuR}. All cases with $\Ynu$ of the type Y$_{1-6}$ or T$_{1-6}$, $\Mr$ of the forms R$_{1,4}$, and degenerate heavy Majorana neutrinos are indicated. For $\Xhnu$ with $(a,b)=\pm(1,-1)$ the matrix $\OCI$ is unconstrained, while for $\Xhnu$ with $(a,b)=\pm(1,1)$ it is restrained to be of the type $\OI$. For completeness, the combinations (R$_1$,Y$_{1-6}$) with $(a,b)=\pm(1,-1)$ have been included in this table. Although not analysed in this paper, those cases are always compatible with data since neither the texture zero nor the CP symmetry lead to low-energy constraints.}
	\label{tablecompdeg}
\end{table}
\begin{table}[t!]
	\centering
	\setlength\extrarowheight{0.7pt}
	\begin{tabular}{cccc}
		\hline\hline
		\multicolumn{4}{c}{Nondegenerate RH neutrinos ($r_N \neq 1$)
		}\\ 
		\hline\hline
		$\Mr$&$\Ynu$&$\OCI$&Compatibility (at least $3\sigma$)\\
		\hline
		\multirow{9}{*}{R$_1$}&T$_{1-6}$&$\mathbf{O}_\text{I-III}$&\xmark\\
		\cline{2-4}
		&\multirow{2}{*}{Y$_1$}&$\OCI_\text{I,III}$&NO, IO\\
		&&$\OII$&\xmark\\
		\cline{2-4}
		&\multirow{2}{*}{Y$_{2,3}$}&$\OCI_\text{I,II}$&NO, IO\\
		&&$\OIII$&IO\\		
		\cline{2-4}
		&\multirow{2}{*}{Y$_4$}&$\OCI_\text{I,II}$&NO, IO\\
		&&$\OIII$&\xmark\\	
		\cline{2-4}
		&\multirow{2}{*}{Y$_{5,6}$}&$\OCI_\text{I,III}$&NO, IO\\
		&&$\OII$&IO\\		
		\hline	
		\multirow{11}{*}{R$_2$}&T$_{1,2}$&$\mathbf{O}_\text{I-III}$&\xmark\\
		\cline{2-4}
		&\multirow{2}{*}{T$_{3,5}$}&$\mathbf{O}_\text{I,III}$&IO; $\xi=\pm1$\\
		&&$\mathbf{O}_\text{II}$&\xmark\\
		\cline{2-4}
		&\multirow{2}{*}{T$_{4,6}$}&$\mathbf{O}_\text{I,II}$&IO; $\xi=\pm1$\\
		&&$\mathbf{O}_\text{III}$&\xmark\\
		\cline{2-4}
		&\multirow{1}{*}{Y$_{1}$}&$\mathbf{O}_\text{I-III}$&\xmark\\
		\cline{2-4}
		&\multirow{1}{*}{Y$_{2,3}$}&$\mathbf{O}_\text{I-III}$&IO; $\xi=\pm1$\\
		\cline{2-4}
		&\multirow{2}{*}{Y$_4$}&$\mathbf{O}_\text{I,II}$&NO, IO; $\xi=\pm1$\\
		&&$\OIII$&\xmark\\		
		\cline{2-4}
		&\multirow{2}{*}{Y$_{5,6}$}&$\mathbf{O}_\text{I,III}$&NO, IO; $\xi=\pm1$\\
		&&$\OII$&IO; $\xi=\pm1$\\
		\hline	
		\multirow{11}{*}{R$_3$}		&\multirow{2}{*}{T$_{1,3}$}&$\mathbf{O}_\text{I,II}$&IO; $\xi=\pm1$\\
		&&$\mathbf{O}_\text{III}$&\xmark\\
		\cline{2-4}
		&\multirow{2}{*}{T$_{2,6}$}&$\mathbf{O}_\text{I,III}$&IO; $\xi=\pm1$\\
		&&$\mathbf{O}_\text{II}$&\xmark\\
		\cline{2-4}
		&T$_{4,5}$&$\mathbf{O}_\text{I-III}$&\xmark\\
		\cline{2-4}
		&\multirow{2}{*}{Y$_1$}&$\mathbf{O}_\text{I,II}$&NO, IO; $\xi=\pm1$\\
		&&$\OIII$&\xmark\\		
		\cline{2-4}
		&\multirow{2}{*}{Y$_{2,3}$}&$\mathbf{O}_\text{I,III}$&NO, IO; $\xi=\pm1$\\
		&&$\OII$&IO; $\xi=\pm1$\\
		\cline{2-4}
		&\multirow{1}{*}{Y$_{4}$}&$\mathbf{O}_\text{I-III}$&\xmark\\
		\cline{2-4}
		&\multirow{1}{*}{Y$_{5,6}$}&$\mathbf{O}_\text{I-III}$&IO; $\xi=\pm1$\\		
		\hline\hline
	\end{tabular}
	\caption{Compatibility summary for the 2RHNSM in presence of the remnant CP symmetry defined in eqs.~\eqref{CPnuL} and~\eqref{CPnuR}. All cases with $\Ynu$ of the type Y$_{1-6}$ or T$_{1-6}$, $\Mr$ of the forms R$_{1-3}$, and nondegenerate heavy Majorana neutrinos are indicated. The compatibility for the R$_{2,3}$ cases is verified for at least one value of $r_N$.}
	\label{tablecomp}
\end{table}

\section{Conclusions}
\label{sec5}

In the framework of the 2RHNSM, we have studied all maximally restricted texture-zero patterns for lepton Yukawa and mass matrices in the presence of the remnant CP symmetry defined in eqs.~\eqref{CPnuL} and~\eqref{CPnuR}. Predictions for the CP-violating phases $\delta$ and $\alpha$, as well as for the effective neutrino mass parameter $m_{\beta\beta}$ (relevant for neutrinoless double beta decay), were obtained and confronted with the present experimental data. In tables~\ref{tablecompdeg} and \ref{tablecomp}, we present a texture compatibility summary for degenerate and nondegenerate RH neutrinos, respectively. 

Our results show that, for degenerate RH neutrinos (i.e., when $\Mr$ of the form R$_1$ or R$_4$), the remnant CP symmetry does not have any impact on the allowed parameter space when that symmetry is such that $(a,b)=\pm(1,-1)$ (see eqs.~\eqref{CPnuL}, \eqref{XnuIONO} and the discussion around eq.~\eqref{ONconst2}). Therefore, in these cases, all constraints stem from the texture zeros imposed on $\Ynu$ and both one- or two-texture-zero $\Ynu$ patterns are allowed for IO (see table~\ref{tablecompdeg}). As explained in section~\ref{sec3}, the predictions for $\delta$, $\alpha$ and $m_{\beta\beta}$ are those presented in ref.~\cite{Barreiros:2018ndn}. On the other hand, for $(a,b)=\pm(1,1)$, the matrix $\OCI$ is constrained to be real, implying additional restrictions on the low-energy parameters. The compatible textures are also indicated in table~\ref{tablecompdeg} for this case.

For nondegenerate RH neutrinos ($\Mr$ of the form R$_{1-3}$), the matrix $\OCI$ is parametrised by a single real parameter when the remnant CP symmetry is imposed (see table~\ref{Rforms}). In these cases:
\begin{itemize}
\item{For the combinations $(\text{R}_{1},\text{Y}_{1-6})$, $(\text{R}_{2},\text{Y}_{4-6})$ or $(\text{R}_{3},\text{Y}_{1-3})$, an additional constraint on the low-energy neutrino parameters arises, besides those coming from the existence of texture zeros in $\Ynu$. This is in contrast to what happens for $(\text{R}_{2},\text{Y}_{1-3})$ and $(\text{R}_{3},\text{Y}_{4-6})$, where the real parameter $z$ is free. With one texture zero in $\Ynu$, the compatible patterns are indicated in table~\ref{tablecomp}. Combining the results obtained for the Y textures, it is then possible to draw conclusions about the compatibility of the cases with two texture zeros in $\Ynu$. A summary of the results for textures T is also given in table~\ref{tablecomp}.}
\item{In general, the restrictions on the low-energy parameters for the viable one-texture-zero $\Ynu$ patterns indicated in table~\ref{tablecomp} are stronger when $\Mr$ is diagonal (see figures~\ref{neutrinoMR1}-\ref{alphambbR2}). This is due to the fact that, with $\Mr$ of the type R$_2$ or R$_3$, $\alpha$ and $m_{\beta\beta}$ are sensitive to the mass ratio $r_N$. In turn, this parameter is also constrained by data, as shown in figure~\ref{neutrinoMR2alpharN}. Notice that, in all cases, and depending on the value of $\alpha$, a nontrivial lower bound on $r_N$ can be set.
}
\item{In the case of two texture zeros in $\Ynu$, only the combinations (R$_2$,T$_{3-6}$) and (R$_3$,T$_{1,2,3,6}$) remain valid for IO. For these cases, the $(\alpha,\delta)$ compatibility regions are more restricted than in the one-texture zero cases, being the ratio $r_N$ fixed by data when $\OCI=\OI$.
}
\item{When compared with the general 2RHNSM (cf. figure~\ref{mbb2RHNSM}), our results indicate that the allowed  parameter space of $m_{\beta\beta}$ is severely constrained in some cases. Actually, a mild improvement of the present experimental bounds on neutrinoless double beta decay rates would disfavour the texture combinations (R$_1$,Y$_{1-6}$), for an IO neutrino mass spectrum and $\OCI=\OI$.
}
\end{itemize}

Our findings show the importance of measuring low-energy neutrino parameters with precision for scrutinising texture-zero patterns in the 2RHNSM, when a remnant CP symmetry is imposed. In this sense, future improvements on the measurement of the phase $\delta$ by neutrino oscillation experiments like T2K~\cite{Abe:2018wpn} and NO$\nu$A~\cite{NOvA:2018gge}, and on neutrinoless double beta decay rates~\cite{DellOro:2016tmg,Vergados:2016hso,Giuliani:neutrino} will further constrain the scenarios analysed in this paper. 

\acknowledgments

This work was partially supported by Funda\c{c}{\~a}o para a Ci{\^e}ncia e a Tecnologia (FCT, Portugal) through the projects CFTP-FCT Unit 777 (UID/FIS/00777/2013), CERN/FIS-PAR/0004/2017 and PTDC/FIS-PAR/29436/2017, which are partly funded through POCTI (FEDER), COMPETE, QREN and EU. The work of D.M.B. is supported by the FCT grant SFRH/BD/137127/2018.


\vspace{0.8cm}

\begin{thebibliography}{99}

%\cite{deSalas:2017kay}
\bibitem{deSalas:2017kay}
P.~F.~de Salas, D.~V.~Forero, C.~A.~Ternes, M.~Tortola and J.~W.~F.~Valle,
%``Status of neutrino oscillations 2018: 3$\sigma$ hint for normal mass ordering and improved CP sensitivity,''
Phys.\ Lett.\ B {\bf 782} (2018) 633.
%DOI:10.1016/j.physletb.2018.06.019
%[arXiv:1708.01186 [hep-ph]].
%%CITATION = DOI:10.1016/j.physletb.2018.06.019;%%
%179 citations counted in INSPIRE as of 23 Nov 2018

%\cite{Esteban:2016qun}
\bibitem{Esteban:2016qun} 
I.~Esteban, M.~C.~Gonzalez-Garcia, M.~Maltoni, I.~Martinez-Soler and T.~Schwetz,
%``Updated fit to three neutrino mixing: exploring the accelerator-reactor complementarity,''
JHEP {\bf 1701} (2017) 087.
%DOI:10.1007/JHEP01(2017)087
%[arXiv:1611.01514 [hep-ph]].
%%CITATION = DOI:10.1007/JHEP01(2017)087;%%

%\cite{Capozzi:2016rtj}
\bibitem{Capozzi:2016rtj} 
F.~Capozzi, E.~Lisi, A.~Marrone, D.~Montanino and A.~Palazzo,
%``Neutrino masses and mixings: Status of known and unknown $3\nu$ parameters,''
Nucl.\ Phys.\ B {\bf 908} (2016) 218.
%DOI:10.1016/j.nuclphysb.2016.02.016
%[arXiv:1601.07777 [hep-ph]].
%%CITATION = DOI:10.1016/j.nuclphysb.2016.02.016;%%

%\cite{Abe:2018wpn}
\bibitem{Abe:2018wpn}
K.~Abe {\it et al.} [T2K Collaboration],
%``Search for CP Violation in Neutrino and Antineutrino Oscillations by the T2K Experiment with $2.2\times10^{21}$ Protons on Target,''
Phys.\ Rev.\ Lett.\  {\bf 121} (2018) 171802.
%DOI:10.1103/PhysRevLett.121.171802
%[arXiv:1807.07891 [hep-ex]].
%%CITATION = DOI:10.1103/PhysRevLett.121.171802;%%
%8 citations counted in INSPIRE as of 23 Nov 2018

%\cite{NOvA:2018gge}
\bibitem{NOvA:2018gge}
M.~A.~Acero {\it et al.} [NOvA Collaboration],
%``New constraints on oscillation parameters from $\nu_e$ appearance and $\nu_\mu$ disappearance in the NOvA experiment,''
Phys.\ Rev.\ D {\bf 98} (2018) 032012.
%DOI:10.1103/PhysRevD.98.032012
%[arXiv:1806.00096 [hep-ex]].
%%CITATION = DOI:10.1103/PhysRevD.98.032012;%%
%9 citations counted in INSPIRE as of 28 Sep 2018

%\cite{DellOro:2016tmg}
\bibitem{DellOro:2016tmg} 
S.~Dell'Oro, S.~Marcocci, M.~Viel and F.~Vissani,
%``Neutrinoless double beta decay: 2015 review,''
Adv.\ High Energy Phys.\  {\bf 2016} (2016) 2162659.
%DOI:10.1155/2016/2162659
%[arXiv:1601.07512 [hep-ph]].
%%CITATION = DOI:10.1155/2016/2162659;%%
%146 citations counted in INSPIRE as of 21 Sep 2018

%\cite{Vergados:2016hso}
\bibitem{Vergados:2016hso} 
J.~D.~Vergados, H.~Ejiri and F.~Šimkovic,
%``Neutrinoless double beta decay and neutrino mass,''
Int.\ J.\ Mod.\ Phys.\ E {\bf 25} (2016) 1630007.
%DOI:10.1142/S0218301316300071
%[arXiv:1612.02924 [hep-ph]].
%%CITATION = DOI:10.1142/S0218301316300071;%%
%61 citations counted in INSPIRE as of 21 Sep 2018

%\cite{Giuliani:neutrino}
\bibitem{Giuliani:neutrino} 
Andrea Giuliani,
``The Mid and Long Term Future of Neutrinoless Double Beta Decay'',
Talk at XXVIII International Conference on Neutrino Physics and Astrophysics, 4-9 June 2018, Heidelberg, Germany, DOI:10.5281/zenodo.1286915.
%URL: https://doi.org/10.5281/zenodo.1286915.

%\cite{Minkowski:1977sc}
\bibitem{Minkowski:1977sc}
P.~Minkowski,
%``$\mu \to e\gamma$ at a Rate of One Out of $10^{9}$ Muon Decays?,''
Phys.\ Lett.\  {\bf 67B} (1977) 421.
%DOI:10.1016/0370-2693(77)90435-X
%%CITATION = DOI:10.1016/0370-2693(77)90435-X;%%
%3258 citations counted in INSPIRE as of 04 Dec 2018

%\cite{GellMann:1980vs}
\bibitem{GellMann:1980vs}
M.~Gell-Mann, P.~Ramond and R.~Slansky,
%``Complex Spinors and Unified Theories,''
Conf.\ Proc.\ C {\bf 790927} (1979) 315.
%[arXiv:1306.4669 [hep-th]].
%%CITATION = ARXIV:1306.4669;%%
%2795 citations counted in INSPIRE as of 04 Dec 2018

%\cite{Yanagida:1979as}
\bibitem{Yanagida:1979as}
T.~Yanagida,
%``Horizontal Symmetry And Masses Of Neutrinos,''
Conf.\ Proc.\ C {\bf 7902131} (1979) 95.
%%CITATION = CONFP,C7902131,95;%%
%1533 citations counted in INSPIRE as of 04 Dec 2018

%\cite{Schechter:1980gr}
\bibitem{Schechter:1980gr}
J.~Schechter and J.~W.~F.~Valle,
%``Neutrino Masses in SU(2) x U(1) Theories,''
Phys.\ Rev.\ D {\bf 22} (1980) 2227.
%DOI:10.1103/PhysRevD.22.2227
%%CITATION = DOI:10.1103/PhysRevD.22.2227;%%
%2421 citations counted in INSPIRE as of 23 Nov 2018

%\cite{Glashow:1979nm}
\bibitem{Glashow:1979nm}
S.~L.~Glashow,
%``The Future of Elementary Particle Physics,''
NATO Sci.\ Ser.\ B {\bf 61} (1980) 687.
%DOI:10.1007/978-1-4684-7197-7_15
%%CITATION = DOI:10.1007/978-1-4684-7197-7_15;%%
%345 citations counted in INSPIRE as of 04 Dec 2018

%\cite{Mohapatra:1979ia}
\bibitem{Mohapatra:1979ia}
R.~N.~Mohapatra and G.~Senjanovic,
%``Neutrino Mass and Spontaneous Parity Nonconservation,''
Phys.\ Rev.\ Lett.\  {\bf 44} (1980) 912.
%DOI:10.1103/PhysRevLett.44.912
%%CITATION = DOI:10.1103/PhysRevLett.44.912;%%
%4828 citations counted in INSPIRE as of 04 Dec 2018

%\cite{Ecker:1981wv}
\bibitem{Ecker:1981wv} 
G.~Ecker, W.~Grimus and W.~Konetschny,
%``Quark Mass Matrices in Left-right Symmetric Gauge Theories,''
Nucl.\ Phys.\ B {\bf 191} (1981) 465.
%DOI:10.1016/0550-3213(81)90309-6
%%CITATION = DOI:10.1016/0550-3213(81)90309-6;%%

%\cite{Ecker:1987qp}
\bibitem{Ecker:1987qp} 
G.~Ecker, W.~Grimus and H.~Neufeld,
%``A Standard Form for Generalized {CP} Transformations,''
J.\ Phys.\ A {\bf 20} (1987) L807.
%DOI:10.1088/0305-4470/20/12/010
%%CITATION = DOI:10.1088/0305-4470/20/12/010;%%

%\cite{Neufeld:1987wa}
\bibitem{Neufeld:1987wa} 
H.~Neufeld, W.~Grimus and G.~Ecker,
%``Generalized {CP} Invariance, Neutral Flavor Conservation and the Structure of the Mixing Matrix,''
Int.\ J.\ Mod.\ Phys.\ A {\bf 3} (1988) 603.
%DOI:10.1142/S0217751X88000254
%%CITATION = DOI:10.1142/S0217751X88000254;%%

%\cite{Grimus:1995zi}
\bibitem{Grimus:1995zi} 
W.~Grimus and M.~N.~Rebelo,
%``Automorphisms in gauge theories and the definition of CP and P,''
Phys.\ Rept.\  {\bf 281} (1997) 239.
%DOI:10.1016/S0370-1573(96)00030-0
%[hep-ph/9506272].
%%CITATION = DOI:10.1016/S0370-1573(96)00030-0;%%

%\cite{Branco:2005em}
\bibitem{Branco:2005em}
G.~C.~Branco, M.~N.~Rebelo and J.~I.~Silva-Marcos,
%``CP-odd invariants in models with several Higgs doublets,''
Phys.\ Lett.\ B {\bf 614} (2005) 187.
%DOI:10.1016/j.physletb.2005.03.075
%[hep-ph/0502118].
%%CITATION = DOI:10.1016/j.physletb.2005.03.075;%%
%83 citations counted in INSPIRE as of 12 Oct 2018

%\cite{Feruglio:2012cw}
\bibitem{Feruglio:2012cw} 
F.~Feruglio, C.~Hagedorn and R.~Ziegler,
%``Lepton Mixing Parameters from Discrete and CP Symmetries,''
JHEP {\bf 1307} (2013) 027.
%DOI:10.1007/JHEP07(2013)027
%[arXiv:1211.5560 [hep-ph]].
%%CITATION = DOI:10.1007/JHEP07(2013)027;%%

%\cite{Holthausen:2012dk}
\bibitem{Holthausen:2012dk} 
M.~Holthausen, M.~Lindner and M.~A.~Schmidt,
%``CP and Discrete Flavour Symmetries,''
JHEP {\bf 1304} (2013) 122.
%DOI:10.1007/JHEP04(2013)122
%[arXiv:1211.6953 [hep-ph]].
%%CITATION = DOI:10.1007/JHEP04(2013)122;%%

%\cite{Girardi:2013sza}
\bibitem{Girardi:2013sza}
I.~Girardi, A.~Meroni, S.~T.~Petcov and M.~Spinrath,
%``Generalised geometrical CP violation in a T' lepton flavour model,''
JHEP {\bf 1402} (2014) 050.
%DOI:10.1007/JHEP02(2014)050
%[arXiv:1312.1966 [hep-ph]].
%%CITATION = DOI:10.1007/JHEP02(2014)050;%%
%40 citations counted in INSPIRE as of 29 Sep 2018

%\cite{Chen:2014tpa}
\bibitem{Chen:2014tpa}
M.~C.~Chen, M.~Fallbacher, K.~T.~Mahanthappa, M.~Ratz and A.~Trautner,
%``CP Violation from Finite Groups,''
Nucl.\ Phys.\ B {\bf 883} (2014) 267.
%DOI:10.1016/j.nuclphysb.2014.03.023
%[arXiv:1402.0507 [hep-ph]].
%%CITATION = DOI:10.1016/j.nuclphysb.2014.03.023;%%
%73 citations counted in INSPIRE as of 10 Oct 2018

%\cite{King:2014rwa}
\bibitem{King:2014rwa}
S.~F.~King and T.~Neder,
%``Lepton mixing predictions including Majorana phases from Δ(6n$^2$) flavour symmetry and generalised CP,''
Phys.\ Lett.\ B {\bf 736} (2014) 308.
%DOI:10.1016/j.physletb.2014.07.043
%[arXiv:1403.1758 [hep-ph]].
%%CITATION = DOI:10.1016/j.physletb.2014.07.043;%%
%52 citations counted in INSPIRE as of 29 Sep 2018

%\cite{Ding:2014ora}
\bibitem{Ding:2014ora}
G.~J.~Ding, S.~F.~King and T.~Neder,
%``Generalised CP and $\Delta(6n^2)$ family symmetry in semi-direct models of leptons,''
JHEP {\bf 1412} (2014) 007.
%DOI:10.1007/JHEP12(2014)007
%[arXiv:1409.8005 [hep-ph]].
%%CITATION = DOI:10.1007/JHEP12(2014)007;%%
%50 citations counted in INSPIRE as of 29 Sep 2018

%\cite{Chen:2014wxa}
\bibitem{Chen:2014wxa} 
P.~Chen, C.~C.~Li and G.~J.~Ding,
%``Lepton Flavor Mixing and CP Symmetry,''
Phys.\ Rev.\ D {\bf 91} (2015) 033003.
%DOI:10.1103/PhysRevD.91.033003
%[arXiv:1412.8352 [hep-ph]].
%%CITATION = DOI:10.1103/PhysRevD.91.033003;%%

%\cite{Everett:2015oka}
\bibitem{Everett:2015oka} 
L.~L.~Everett, T.~Garon and A.~J.~Stuart,
%``A Bottom-Up Approach to Lepton Flavor and CP Symmetries,''
JHEP {\bf 1504} (2015) 069.
%DOI:10.1007/JHEP04(2015)069
%[arXiv:1501.04336 [hep-ph]].
%%CITATION = DOI:10.1007/JHEP04(2015)069;%%

%\cite{Branco:2015hea}
\bibitem{Branco:2015hea} 
G.~C.~Branco, I.~de Medeiros Varzielas and S.~F.~King,
%``Invariant approach to CP in family symmetry models,''
Phys.\ Rev.\ D {\bf 92} (2015) 036007.
%DOI:10.1103/PhysRevD.92.036007
%[arXiv:1502.03105 [hep-ph]].
%%CITATION = DOI:10.1103/PhysRevD.92.036007;%%

%\cite{Turner:2015uta}
\bibitem{Turner:2015uta}
P.~Ballett, S.~Pascoli and J.~Turner,
%``Mixing angle and phase correlations from A5 with generalized CP and their prospects for discovery,''
Phys.\ Rev.\ D {\bf 92} (2015) 093008.
%DOI:10.1103/PhysRevD.92.093008
%[arXiv:1503.07543 [hep-ph]].
%%CITATION = DOI:10.1103/PhysRevD.92.093008;%%
%48 citations counted in INSPIRE as of 29 Sep 2018


%\cite{Chen:2015nha}
\bibitem{Chen:2015nha} 
P.~Chen, C.~Y.~Yao and G.~J.~Ding,
%``Neutrino Mixing from CP Symmetry,''
Phys.\ Rev.\ D {\bf 92} (2015) 073002.
%DOI:10.1103/PhysRevD.92.073002
%[arXiv:1507.03419 [hep-ph]].
%%CITATION = DOI:10.1103/PhysRevD.92.073002;%%

%\cite{Girardi:2015rwa}
\bibitem{Girardi:2015rwa}
I.~Girardi, S.~T.~Petcov, A.~J.~Stuart and A.~V.~Titov,
%``Leptonic Dirac CP Violation Predictions from Residual Discrete Symmetries,''
Nucl.\ Phys.\ B {\bf 902} (2016) 1.
%DOI:10.1016/j.nuclphysb.2015.10.020
%[arXiv:1509.02502 [hep-ph]].
%%CITATION = DOI:10.1016/j.nuclphysb.2015.10.020;%%
%33 citations counted in INSPIRE as of 10 Oct 2018

%\cite{Chen:2016ica}
\bibitem{Chen:2016ica} 
P.~Chen, G.~J.~Ding, F.~Gonzalez-Canales and J.~W.~F.~Valle,
%``Classifying CP transformations according to their texture zeros: theory and implications,''
Phys.\ Rev.\ D {\bf 94} (2016) 033002.
%DOI:10.1103/PhysRevD.94.033002
%[arXiv:1604.03510 [hep-ph]].
%%CITATION = DOI:10.1103/PhysRevD.94.033002;%%

%\cite{Penedo:2017vtf}
\bibitem{Penedo:2017vtf}
J.~T.~Penedo, S.~T.~Petcov and A.~V.~Titov,
%``Neutrino mixing and leptonic CP violation from S$_{4}$ flavour and generalised CP symmetries,''
JHEP {\bf 1712} (2017) 022.
%DOI:10.1007/JHEP12(2017)022
%[arXiv:1705.00309 [hep-ph]].
%%CITATION = DOI:10.1007/JHEP12(2017)022;%%
%14 citations counted in INSPIRE as of 29 Sep 2018

%\cite{Ivanov:2017bdx}
\bibitem{Ivanov:2017bdx}
I.~P.~Ivanov,
%``Radiative neutrino masses from order-4 CP symmetry,''
JHEP {\bf 1802} (2018) 025.
%DOI:10.1007/JHEP02(2018)025
%[arXiv:1712.02101 [hep-ph]].
%%CITATION = DOI:10.1007/JHEP02(2018)025;%%
%2 citations counted in INSPIRE as of 29 Sep 2018

%\cite{Samanta:2017kce}
\bibitem{Samanta:2017kce}
R.~Samanta, P.~Roy and A.~Ghosal,
%``Consequences of minimal seesaw with complex $\mu\tau$ antisymmetry of neutrinos,''
JHEP {\bf 1806} (2018) 085.
%DOI:10.1007/JHEP06(2018)085
%[arXiv:1712.06555 [hep-ph]].
%%CITATION = DOI:10.1007/JHEP06(2018)085;%%
%11 citations counted in INSPIRE as of 17 Oct 2018

%\cite{Chen:2018lsv}
\bibitem{Chen:2018lsv}
P.~Chen, S.~Centelles Chuliá, G.~J.~Ding, R.~Srivastava and J.~W.~F.~Valle,
%``Neutrino Predictions from Generalized CP Symmetries of Charged Leptons,''
JHEP {\bf 1807} (2018) 077.
%DOI:10.1007/JHEP07(2018)077
%[arXiv:1802.04275 [hep-ph]].
%%CITATION = DOI:10.1007/JHEP07(2018)077;%%
%3 citations counted in INSPIRE as of 23 Nov 2018

%\cite{Li:2017zmk}
\bibitem{Li:2017zmk} 
C.~C.~Li and G.~J.~Ding,
%``Implications of remnant CP symmetry for leptogenesis in a model with two right-handed neutrinos,''
Phys.\ Rev.\ D {\bf 96} (2017) 075005.
%DOI:10.1103/PhysRevD.96.075005
%[arXiv:1701.08508 [hep-ph]].
%%CITATION = DOI:10.1103/PhysRevD.96.075005;%%


%\cite{Casas:2001sr}
\bibitem{Casas:2001sr} 
J.~A.~Casas and A.~Ibarra,
%``Oscillating neutrinos and muon ---> e, gamma,''
Nucl.\ Phys.\ B {\bf 618} (2001) 171.
%DOI:10.1016/S0550-3213(01)00475-8
%[hep-ph/0103065].
%%CITATION = DOI:10.1016/S0550-3213(01)00475-8;%%

\bibitem{Grimus:2004hf}
W.~Grimus, A.~S.~Joshipura, L.~Lavoura and M.~Tanimoto,
%``Symmetry realization of texture zeros,''
Eur.\ Phys.\ J.\ C {\bf 36} (2004) 227.
%DOI:10.1140/epjc/s2004-01896-y,
%[hep-ph/0405016].
%%CITATION = HEP-PH/0405016;%%
%
%\cite{Dighe:2009xj}
\bibitem{Dighe:2009xj}
A.~Dighe and N.~Sahu,
%``Texture zeroes and discrete flavor symmetries in light and heavy Majorana neutrino mass matrices: a bottom-up approach,''
arXiv:0812.0695 [hep-ph].
%%CITATION = ARXIV:0812.0695;%%
%19 citations counted in INSPIRE as of 29 Sep 2018
%
%\cite{Adhikary:2009kz}
\bibitem{Adhikary:2009kz}
B.~Adhikary, A.~Ghosal and P.~Roy,
%``mu tau symmetry, tribimaximal mixing and four zero neutrino Yukawa textures,''
JHEP {\bf 0910} (2009) 040.
%DOI:10.1088/1126-6708/2009/10/040
%[arXiv:0908.2686 [hep-ph]].
%%CITATION = DOI:10.1088/1126-6708/2009/10/040;%%
%42 citations counted in INSPIRE as of 29 Sep 2018
%
%\cite{Dev:2011jc}
\bibitem{Dev:2011jc}
S.~Dev, S.~Gupta and R.~R.~Gautam,
%``Zero Textures of the Neutrino Mass Matrix from Cyclic Family Symmetry,''
Phys.\ Lett.\ B {\bf 701} (2011) 605.
%DOI:10.1016/j.physletb.2011.06.046
%[arXiv:1106.3451 [hep-ph]].
%%CITATION = DOI:10.1016/j.physletb.2011.06.046;%%
%23 citations counted in INSPIRE as of 29 Sep 2018
%
%\cite{Felipe:2014vka}
\bibitem{Felipe:2014vka} 
R.~Gonz\'{a}lez Felipe and H.~Ser\^{o}dio,
%``Abelian realization of phenomenological two-zero neutrino textures,''
Nucl.\ Phys.\ B {\bf 886} (2014) 75.
%DOI:10.1016/j.nuclphysb.2014.06.015
%[arXiv:1405.4263 [hep-ph]].
%%CITATION = DOI:10.1016/j.nuclphysb.2014.06.015;%%
%
%\cite{Cebola:2015dwa}
\bibitem{Cebola:2015dwa} 
L.~M.~Cebola, D.~Emmanuel-Costa and R.~G.~Felipe,
%``Confronting predictive texture zeros in lepton mass matrices with current data,''
Phys.\ Rev.\ D {\bf 92} (2015) 025005.
%DOI:10.1103/PhysRevD.92.025005.
%[arXiv:1504.06594 [hep-ph]].
%%CITATION = DOI:10.1103/PhysRevD.92.025005;%%
%
%\cite{Samanta:2015oqa}
\bibitem{Samanta:2015oqa}
R.~Samanta and A.~Ghosal,
%``Probing maximal zero textures with broken cyclic symmetry in inverse seesaw,''
Nucl.\ Phys.\ B {\bf 911} (2016) 846.
%DOI:10.1016/j.nuclphysb.2016.08.036
%[arXiv:1507.02582 [hep-ph]].
%%CITATION = DOI:10.1016/j.nuclphysb.2016.08.036;%%
%6 citations counted in INSPIRE as of 29 Sep 2018
%
%\cite{Kobayashi:2018zpq}
\bibitem{Kobayashi:2018zpq}
T.~Kobayashi, T.~Nomura and H.~Okada,
%``Predictive neutrino mass textures with origin of flavor symmetries,''
Phys.\ Rev.\ D {\bf 98} (2018) 055025.
%DOI:10.1103/PhysRevD.98.055025
%[arXiv:1805.07101 [hep-ph]].
%%CITATION = DOI:10.1103/PhysRevD.98.055025;%%
%2 citations counted in INSPIRE as of 23 Nov 2018

%\cite{Rahat:2018sgs}
\bibitem{Rahat:2018sgs}
M.~H.~Rahat, P.~Ramond and B.~Xu,
%``Asymmetric tribimaximal texture,''
Phys.\ Rev.\ D {\bf 98} (2018) 055030.
%DOI:10.1103/PhysRevD.98.055030
%[arXiv:1805.10684 [hep-ph]].
%%CITATION = DOI:10.1103/PhysRevD.98.055030;%%
%1 citations counted in INSPIRE as of 17 Oct 2018

%\cite{Nath:2018xih}
\bibitem{Nath:2018xih}
N.~Nath,
%``$ \mu-\tau $ Reflection Symmetry and Its Explicit Breaking for Leptogenesis in a Minimal Seesaw Model,''
arXiv:1808.05062 [hep-ph].
%%CITATION = ARXIV:1808.05062;%%
%3 citations counted in INSPIRE as of 17 Oct 2018

%\cite{Frampton:2002qc}
\bibitem{Frampton:2002qc} 
P.~H.~Frampton, S.~L.~Glashow and T.~Yanagida,
%``Cosmological sign of neutrino CP violation,''
Phys.\ Lett.\ B {\bf 548} (2002) 119.
%DOI:10.1016/S0370-2693(02)02853-8.
%[hep-ph/0208157].
%%CITATION = DOI:10.1016/S0370-2693(02)02853-8;%%
%

%\cite{zerosNeutrino}
\bibitem{Ibarra:2003up} 
A.~Ibarra and G.~G.~Ross,
%``Neutrino phenomenology: The Case of two right-handed neutrinos,''
Phys.\ Lett.\ B {\bf 591} (2004) 285.
%DOI:10.1016/j.physletb.2004.04.037.
%[hep-ph/0312138].
%%CITATION = DOI:10.1016/j.physletb.2004.04.037;%%
%
%\cite{Harigaya:2012bw}
\bibitem{Harigaya:2012bw} 
K.~Harigaya, M.~Ibe and T.~T.~Yanagida,
%``Seesaw Mechanism with Occam's Razor,''
Phys.\ Rev.\ D {\bf 86} (2012) 013002.
%DOI:10.1103/PhysRevD.86.013002.
%[arXiv:1205.2198 [hep-ph]].
%%CITATION = DOI:10.1103/PhysRevD.86.013002;%%
%
%\cite{Rink:2016vvl}
\bibitem{Rink:2016vvl} 
T.~Rink and K.~Schmitz,
%``Perturbed Yukawa Textures in the Minimal Seesaw Model,''
JHEP {\bf 1703} (2017) 158.
%[arXiv:1611.05857 [hep-ph]].
%DOI:10.1007/JHEP03(2017)158.
%%CITATION = DOI:10.1007/JHEP03(2017)158;%%
%
%\cite{Shimizu:2017fgu}
\bibitem{Shimizu:2017fgu}
Y.~Shimizu, K.~Takagi and M.~Tanimoto,
%``Towards the minimal seesaw model via CP violation of neutrinos,''
JHEP {\bf 1711} (2017) 201.
%[arXiv:1709.02136 [hep-ph]].
%DOI:10.1007/JHEP11(2017)201
%%CITATION = DOI:10.1007/JHEP11(2017)201;%%
%7 citations counted in INSPIRE as of 23 Nov 2018

%\cite{Barreiros:2018ndn}
\bibitem{Barreiros:2018ndn} 
D.~M.~Barreiros, R.~G.~Felipe and F.~R.~Joaquim,
%``Minimal type-I seesaw model with maximally restricted texture zeros,''
Phys.\ Rev.\ D {\bf 97} (2018) 115016.
%[arXiv:1802.04563 [hep-ph]].
%DOI:10.1103/PhysRevD.97.115016
%%CITATION = DOI:10.1103/PhysRevD.97.115016;%%

%\cite{Alcaide:2018vni}
\bibitem{Alcaide:2018vni}
J.~Alcaide, J.~Salvado and A.~Santamaria,
%``Fitting flavour symmetries: the case of two-zero neutrino mass textures,''
JHEP {\bf 1807} (2018) 164.
%DOI:10.1007/JHEP07(2018)164
%[arXiv:1806.06785 [hep-ph]].
%%CITATION = DOI:10.1007/JHEP07(2018)164;%%
%2 citations counted in INSPIRE as of 29 Sep 2018

%\cite{Branco:2011zb}
\bibitem{Branco:2011zb} 
G.~C.~Branco, R.~G.~Felipe and F.~R.~Joaquim,
%``Leptonic CP Violation,''
Rev.\ Mod.\ Phys.\  {\bf 84} (2012) 515.
%[arXiv:1111.5332 [hep-ph]].
%%CITATION = ARXIV:1111.5332;%%

\bibitem{Rodejohann:2011vc}
W.~Rodejohann and J.~W.~F.~Valle,
%``Symmetrical Parametrizations of the Lepton Mixing Matrix,''
Phys.\ Rev.\ D {\bf 84} (2011) 073011.
%doi:10.1103/PhysRevD.84.073011
%[arXiv:1108.3484 [hep-ph]].
%%CITATION = doi:10.1103/PhysRevD.84.073011;%%
%51 citations counted in INSPIRE as of 07 Jan 2019

%\cite{Heeck:2012fw}
\bibitem{Heeck:2012fw} 
J.~Heeck,
%``Seesaw parametrization for n right-handed neutrinos,''
Phys.\ Rev.\ D {\bf 86} (2012) 093023.
%[arXiv:1207.5521 [hep-ph]].
%DOI:10.1103/PhysRevD.86.093023
%%CITATION = DOI:10.1103/PhysRevD.86.093023;%%
%14 citations counted in INSPIRE as of 12 Sep 2018

%\cite{Ibarra:2011xn}
\bibitem{Ibarra:2011xn}
A.~Ibarra, E.~Molinaro and S.~T.~Petcov,
%``Low Energy Signatures of the TeV Scale See-Saw Mechanism,''
Phys.\ Rev.\ D {\bf 84} (2011) 013005.
%[arXiv:1103.6217 [hep-ph]].
%DOI:10.1103/PhysRevD.84.013005
%%CITATION = DOI:10.1103/PhysRevD.84.013005;%%
%108 citations counted in INSPIRE as of 10 Oct 2018

%\cite{Alduino:2017ehq}
\bibitem{Alduino:2017ehq}
C.~Alduino {\it et al.} [CUORE Collaboration],
%``First Results from CUORE: A Search for Lepton Number Violation via $0\nu\beta\beta$ Decay of $^{130}$Te,''
Phys.\ Rev.\ Lett.\  {\bf 120} (2018) 132501.
%[arXiv:1710.07988 [nucl-ex]].
%DOI:10.1103/PhysRevLett.120.132501
%%CITATION = DOI:10.1103/PhysRevLett.120.132501;%%
%53 citations counted in INSPIRE as of 25 Sep 2018

%\cite{Asakura:2015ajs}
\bibitem{Asakura:2015ajs}
K.~Asakura {\it et al.} [KamLAND-Zen Collaboration],
%``Search for double-beta decay of $^{136}$Xe to excited states of $^{136}$Ba with the KamLAND-Zen experiment,''
Nucl.\ Phys.\ A {\bf 946} (2016) 171.
%[arXiv:1509.03724 [hep-ex]].
%DOI:10.1016/j.nuclphysa.2015.11.011
%%CITATION = DOI:10.1016/j.nuclphysa.2015.11.011;%%
%15 citations counted in INSPIRE as of 25 Sep 2018

\end{thebibliography}
\end{document}